\UseRawInputEncoding

\documentclass[]{raa}            

\usepackage{graphicx,times}             
\usepackage{natbib}
\usepackage{amssymb,amsmath}
\bibpunct{(}{)}{;}{a}{}{,}

\usepackage{longtable}
\usepackage{grffile}
\usepackage{ulem}

\usepackage[pagebackref=true]{hyperref}

\newcommand{\kpc}{\ensuremath{\,{\rm kpc}}}

\begin{document}

  \title{Classifying globular clusters and applying them to estimate the mass of the Milky Way}


   \volnopage{Vol.0 (20xx) No.0, 000--000}      
   \setcounter{page}{1}          

   \author{GuangChen Sun
      \inst{1,2,3,4}
   \and Yougang Wang \thanks{E-mail:wangyg@bao.ac.cn}
      \inst{3,4}
   \and Chao Liu
      \inst{3,4}
   \and Richard J.  Long
      \inst{5,6}
   \and Xuelei Chen
      \inst{3,4,7,8}
   \and Qi Gao  \thanks{E-mail:fbc1980@163.com}
      \inst{1,2}
   }

   \institute{Department of Physics, College of Science, Tibet University, Lhasa 850000, P. R. China \\
        \and
             Key Laboratory of Cosmic Rays (Tibet University), Ministry of Education, Lhasa 850000, P. R. China\\
        \and
             National Astronomical Observatories, Chinese Academy of Sciences, Beijing 100101, China\\
        \and
            School of Astronomy and Space Science, University of Chinese Academy of Sciences, Beijing 100049, China\\
        \and
            Department of Astronomy, Tsinghua University, Beijing, 100084, China\\
        \and
            Jodrell Bank Centre for Astrophysics, Department of Physics and Astronomy, The University of Manchester, Oxford Road, Manchester M13 9PL, UK\\
        \and
            Department of Physics, College of Sciences, Northeastern University, Shenyang 110819, China\\
        \and
            Center of High Energy Physics, Peking University, Beijing 100871, China \\
\vs\no
   {\small Received 20xx month day; accepted 20xx month day}}

\abstract{We combine the kinematics of 159 globular clusters (GCs) provided by the Gaia Early Data Release 3 (EDR3) with other observational data to classify the GCs, and to estimate the mass of the Milky Way (MW). We use the age-metallicity relation, integrals of motion, action space and the GC orbits to identify the GCs as either formed in-situ (Bulge and Disk) or ex situ (via accretion).
We find that $45.3\%$ have formed in situ, $38.4\%$ may be related to known merger events: Gaia-Sausage-Enceladus, the Sagittarius dwarf galaxy, the Helmi streams, the Sequoia galaxy, and the Kraken galaxy. We also further identify three new sub-structures associated with the Gaia-Sausage-Enceladus.  The remaining $16.3\%$ of GCs are unrelated to the known mergers and thought to be from small accretion events.  We select 46 GCs which have radii $8.0<r<37.3$ kpc and obtain the anisotropy parameter $\beta=0.315_{-0.049}^{+0.055}$, which is  lower than the recent result using the sample of GCs in Gaia Data Release 2, but still in agreement with it by considering the error bar. By using the same sample, we obtain the MW mass inside the outermost GC as $M(<37.3 \kpc)=0.423_{-0.02}^{+0.02}\times10^{12}M_{\sun}$, and the corresponding $M_{200}=1.11_{-0.18}^{+0.25}\times10^{12}M_{\sun}$. The estimated mass is consistent with the results in many recent studies. We also find that the estimated $\beta$ and mass depend on the selected sample of GCs. However, it is difficult to determine whether a GC fully traces the potential of the MW.
\keywords{Galaxy: kinematics and dynamics -- Galaxy: fundamental parameters -- Galaxy: structure -- Galaxy: halo -- dark matter -- globular clusters: general
}
}


   \maketitle

%
%
\section{Introduction}
In the cold dark matter scenario, the visible part of a galaxy is usually embedded in a dark matter halo, and the dark matter halos grow in mass and size via accretion and merger processes ~\citep{white1978}. Accurate determination of the halo mass from luminous tracers is crucial for understanding structure formation and distinguishing between different dark matter models ~\citep[e.g.][]{Bose2018}.

Compared with other distant galaxies, the Milky Way (MW) has the most abundant observational data, and many different tracers, such as bright stars, globular clusters (GCs), HI gas, satellite galaxies and tidal streams. All of these can be used to estimate the mass of the MW. However,
our location within the MW increases the complexity of the mass calculations, and each method of mass estimation has its own systematic error. As shown in Fig. 1 in \cite{Wang2020}, there is a large scatter in the current mass estimates of the MW.

The accuracy of the MW mass estimates can be  improved in two ways. One is to employ more realistic models. For example,  simulations show that dark matter halos are triaxial \citep{Jing2002}, and the spherical assumption brings in a systematic bias \citep{wang2018}. In more realistic models the MW halo is taken as triaxial, and more sophisticated methods such as the Schwarzschild \citep{SCHWARZSCHILDM1979}  or the made-to-measure method \citep{syer1996} can be used. Another often adopted assumption is that the MW is in dynamical equilibrium. However, spiral arms and the warp in the MW are typically non-equilibrium features. Moreover, if the potential of MW slowly evolves with time due to accretion, the system will violate the dynamical equilibrium assumption.  Action-angle space calculations are able to cope with these non-equilibria \citep[e.g.][]{wang2017, Vasiliev2019}.
The other way of improvement is of course to use higher quality data with lower errors in velocity and position. Thanks to the Gaia mission \citep{gaia2018a}, a larger number of accurate proper motion (PM) data has become available now, enabling the use of data with complete phase space information for many objects in the MW.

GCs are some of the oldest objects in the MW \citep[][]{Mar2009} and are bright enough to be easily observed. They are abundant in the MW inner halo, can serve as tracer of the potential and be used to determine the MW mass in this region. There are extensive studies with GCs to constrain the mass of the MW \citep[][]{battaglia2005, Watkins2010, Eadie2016, Sohn2018}.
The accuracy of the MW mass estimation depends on having complete phase space data for each GC. With the PMs of GCs available from Gaia Data Release 2 (Gaia DR2, ~\citealt{GaiaCollaboration2018c}), \cite{Watkins2019} and ~\cite{Eadie2019} improved the MW mass estimates using the GCs, and both of them found that the estimated mass of the MW depends strongly on the selection of the GC sample.     

Recently, the Gaia mission published its Early Data Release 3 (Gaia EDR3) \citep[][]{GaiaEDR32020}. Compared with DR2, the PM precisions are improved by a factor of two in EDR3. Based on Gaia EDR3, ~\cite{Vasiliev2021} measured the mean parallaxes and PMs for 170 GCs, and determined the PM dispersion distribution for more than 100 GCs. As the quantity and quality of the GC data increase, more accurate measurements of the MW mass are expected. The aim of this paper is to improve the mass estimation by using the latest GC sample with complete phase space information.



The paper is structured as follows. In Section~\ref{sec:data}, we describe the data used and the selection method for the halo GC sample. We combine existing observations to explore the spatial motions of our target GCs. In Section~\ref{sec:selection of gcs}, we present our classification results of the GCs.
In Section~\ref{sec:method}, we describe our method of the mass estimation. In Section~\ref{sec:mass of the mw}, we estimate the mass of the MW. Summary and discussion are provided in Section~\ref{sec:conclusions}.

\section{Data}
\label{sec:data}

The PM data ($\mu_\alpha$, $\mu_\delta$) \footnote{Where $\mu_\alpha\equiv (\rm{d}\alpha/\rm{d}t) \cos{\delta}$, and $\mu_\delta \equiv \rm{d}\delta/\rm{d}t$, here $\alpha$ is the right ascension, and $\delta$ is the declination.} we considered are from \citet{Vasiliev2021}, which is based on Gaia EDR3 and includes 170 GCs. Here, $\mu_\alpha$ and $\mu_\delta$ are the PM along the right ascension and declination, respectively. Nine of the 170 GCs do not have  line-of-sight (LOS) velocity information. Since we need six-dimensional phase-space information for each GC, these are removed from our sample. 

The distances to the Sun ($d$) for the GCs are provided by \cite{Baumgardt_2021}. They determined the mean distances to the GCs using a combination of Gaia EDR3 data and distances in published literature.
Note that four GCs are missing from this dataset.  ESO 93-SC8 and BH 176 (ESO 224-8) are thought to be disk GCs \citep{Massari2019, 2021RAA....21..173B}, while Munoz 1 and Segue 3 may well be old open clusters \citep{Vasiliev2021}. For consistency of the data, we removed them, but we referenced existing distance data to evaluate these four GCs and confirmed that they would not have a significant impact on our results.  Therefore, the number of GCs we used from ~\cite{Vasiliev2021} is 157. In addition, two further GCs with PMs, 2MASS-GC01 and 2MASS-GC02, come from \cite{Baumgardt2018}. Since \cite{Vasiliev2021} have shown that there is no obvious systematic error between their PMs and those in \cite{Baumgardt2018}, we have added these two GCs to our sample. Overall, the total number of GCs in our final data set is 159.

\begin{figure}

	\includegraphics[width= \columnwidth]{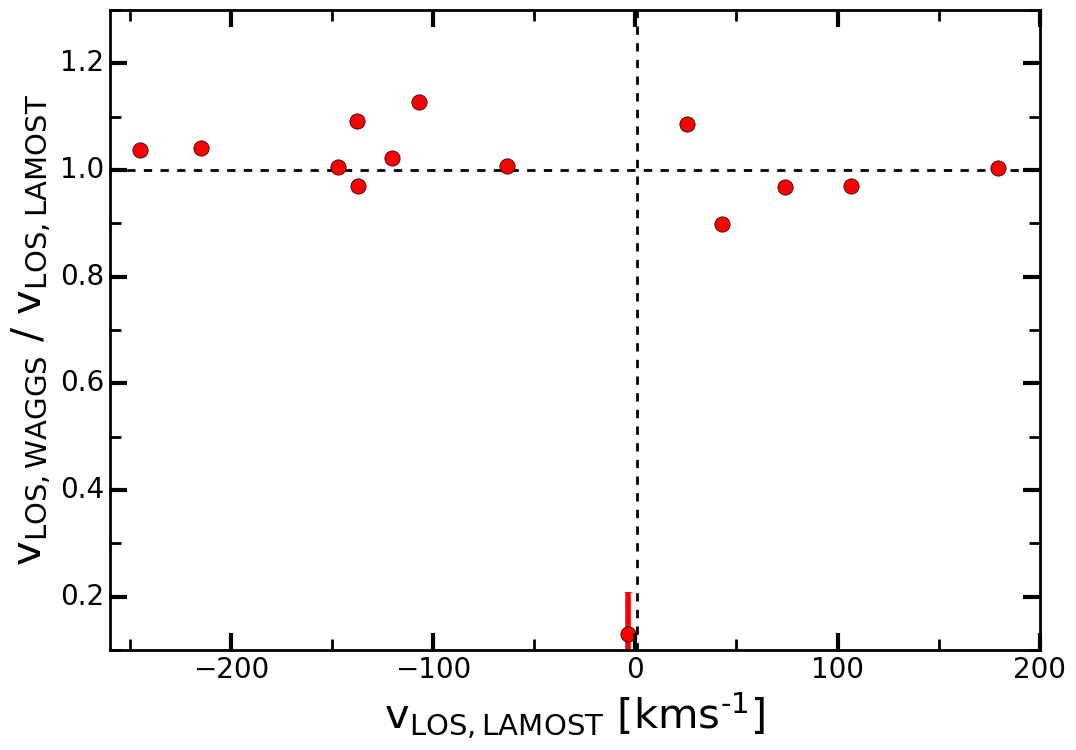}
    \caption{Comparison of the LOS velocities from the WAGGS data with those from LAMOST DR8. The black horizontal line indicates the positions of two data samples are perfectly consistent. Here $v_{\rm LOS,WAGGS}$ and $v_{\rm LOS,LAMOST}$ represent LOS velocities from WAGGS and LAMOST observations, respectively.}
    \label{fig:lamost}
\end{figure}

The LOS velocities $v_{\rm LOS}$ we used are from the WAGGS survey \citep[][]{Dalgleish2020}. To check the accuracy of the LOS velocities, we compared the LOS velocities from the WAGGS data with those from LAMOST Data Release 8 (DR8). The average ratio of the WAGGS LOS velocities to the LAMOST LOS velocities is ${v_{\rm LOS,WAGGS}}/{v_{\rm LOS,LAMOST}}=1.020$. There are 14 GCs in our sample that have both LOS velocity values. As affirmed in Fig.~\ref{fig:lamost}, the LOS velocities from the two data samples are consistent with each other for 13 of the GCs. For the one remaining GC, the ratio of $v_{\rm LOS,WAGGS}/v_{\rm LOS,LAMOST}$ deviates from 1 significantly because the absolute value of $v_{\rm LOS,LAMOST}$ is close to zero. Since the LOS velocity errors in the WAGGS data are smaller than those in the LAMOST data, we take the LOS velocities from the WAGGS data.

Most of the GC metallicities are from \cite{Harris2010}. For some GCs, there is no metallicity information in \cite{Harris2010}. We then supplement the metallicity information for these GCs from other references, such as  BH 176 in \cite{sharina2014}, FSR 1735 in \cite{bello2016}, FSR 1758 in \cite{Barb2019}, 2MASS-GC01 in \cite{borissova2009}, VVV-CL001 in \cite{Fern_ndez_2021}, Mercer 5 in \cite{Pe_aloza_2015}, Segue 3 in \cite{Hughes_2017}, Laevens 3 in \cite{Longeard2019}, ESO 93-SC08 in \cite{bica1999}, and ESO 452-SC11 (1636-283) in ~\cite{Simpson2017}). We list these in Table~\ref{tab:A1}.

The distance of the Sun to the Galactic Center is set to be $R_{\sun}=8.178\pm0.035$\,kpc \citep[][]{GravityCollaboration2019}, with height above the equatorial plane $z_{\sun}=20.8\pm0.3$\,pc \citep{zsun2019}. The velocity of the Sun in the Galactic rest frame is taken to be $v_{\sun}=(U_{\sun}, V_{\sun}, W_{\sun})=(11.10\pm1.75, 12.24\pm2.47, 7.25\pm0.87)$\,km\,s$^{-1}$ \citep[][]{schonrich2010}, where $U_{\sun}$ is the velocity toward the Galactic Center, $V_{\sun}$ is the velocity in the direction of the rotation of the MW, and $W_{\sun}$ that toward the Galactic North Pole.

In order to explore the motion of our selected GCs, we calculate the orbit for each GC. Orbit integration is carried out with the \texttt{AGAMA} code \citep{vasiliev2018}. The adopted potential of the MW is the same as in \citet[hereafter Mc17]{McMillan2017}. In Mc17, the MW is decomposed into an axisymmetric bulge, thin and thick stellar disks, two gas disks and a spherical dark matter halo. 

In observations, there are errors in the PMs ($\mu_\alpha$ and $\mu_\delta$),  LOS velocities ($v_{\rm LOS}$) and distance ($d$) measurements, which will produce the uncertainties in orbit initial conditions and thus the orbit structure. We assume that all observational uncertainties are following a Gaussian distribution. For each GC, we randomly produce 2000 initial conditions.


We use the \texttt{Astropy} library \citep{astropy2013, astropy2018} to convert ($\alpha$, $\delta$, $d$, $v_{\rm los}$, $\mu_\alpha$, $\mu_\delta$) to a Cartesian coordinate system with the Galaxy Center as the origin $(x,y,z,v_x,v_y,v_z)$. Here $x$ is the direction from the Sun to the Galactic Center, $y$ is in the direction of the Galactic rotation and $z$ is the direction pointing toward the North Pole. where $v_x$, $v_y$ and $v_z$ are the velocities in the Galactic rest frame. 

The orbits for all GCs are calculated using \texttt{AGAMA} code \citep{vasiliev2018}.
We also computed the GC orbital parameters apocenter (Apo), pericenter (Peri), maximum height from the plane ($z_{\rm max}$) and eccentricity (ecc) of the orbit given by

\begin{equation}
 \mathrm{ecc}=\frac{\mathrm{Apo-Peri}}{\mathrm{Apo+Peri}}
	\label{eq:ecc}
\end{equation}

In addition, we calculated the actions of the orbits in a cylindrical coordinate system ($J_R$, $J_z$, $J_\phi$), where $J_\phi \equiv L_z = x \times v_y - y \times v_x$ with sense opposite to the Sun's motion. $L_z$ is the $z$-component of the angular momentum $L$ in Cartesian coordinates. In addition, the radial and vertical actions are $J_R$ and $J_z$, respectively. We give the GC orbital motion information in Table~\ref{tab:A2}.

\section{Classification of globular clusters}
\label{sec:selection of gcs}
As shown in \cite{Watkins2019} and ~\cite{Eadie2019}, the mass estimation of the MW depends strongly  on the selected sample of the GCs. To understand why this happens and obtain a good sample of GCs for use, it is important to understand the origin of each GC. Nearly $40\%$ of GCs have an ex situ origin~\citep{kruijssen2018}, and different studies can give different origins even for the same GC~\citep[e.g.][]{2022arXiv}. Therefore, it is necessary to classify the origin of each GC carefully. 


Here we divide the sample GCs into in situ and ex situ groups, and determine the progenitors of the latter samples. We mainly follow the approach of \citet{Massari2019}, \citet[hereafter BB21]{2021RAA....21..173B} and \citet{Duncan2020}. We describe our methods in detail in subsequent subsections.

We divide the ex situ GCs into Sagittarius dwarf \citep[][Sgr]{ibata1994, law2010b, bellazzini2020}, Sequoia galaxy \citep[][Seq]{myeong2019}, Helmi Streams \citep[][Hel]{HELMI1999, koppelman2019}, Gaia-Sausage-Enceladus \citep[][GSE]{belokurov2018, helmi2018}, Kraken \footnote{Basically corresponds to the low-energy group in \citet{Massari2019}. In order to avoid confusion, Kraken in the following parts of this paper is equivalent to the low-energy GCs in \citet{Massari2019}.} \citep[][]{kruijssen2018, Duncan2020, Kruijssen2020} and potential accretion (Pot) GCs. Here the Pot GCs are those which cannot be associated with a known merger event, and some of them are consistent with those of the high-energy group in \cite{Massari2019}.
The Pot GCs may be small objects which have merged with the MW in the past, for example, Pyxis \citep[][]{Fritz_2017}.  Most Pot GCs are in high energy areas. We list the grouping of GCs  in Table~\ref{tab:table2}.

\begin{table*}
	\centering
	\renewcommand{\arraystretch}{1}
    \setlength\tabcolsep{5pt}
	\caption{Progenitors and their associated GCs by Name and Numbers.} 
	\label{tab:table2}
	\begin{tabular}{lcc}
		\hline
		Progenitor & GCs & Number\\
		\hline
		Bulge & BH 229 (HP 1), Djorg 2 (ESO 456-SC38), ESO452-SC11 (1636-283), Liller 1, NGC 6093 (M80), NGC 6144, & 38\\ 
		& NGC 6171 (M107), NGC 6266 (M62), NGC 6293, NGC 6316, NGC 6325, NGC 6342, NGC 6355, NGC 6380 (Ton 1), \\
		& NGC 6388, NGC 6401, NGC 6440, NGC 6453, NGC 6517, NGC 6522, NGC 6528, NGC 6540 (Djorg 3), NGC 6558, \\
		& NGC 6624, NGC 6626 (M28), NGC 6637 (M69), NGC 6638, NGC 6642, NGC 6652, NGC 6717 (Pal 9), NGC 6723, \\
		& Pal 6, Terzan 1 (HP 2), Terzan 2 (HP 3), Terzan 4 (HP 4), Terzan 5 (Ter 11), Terzan 6 (HP 5), Terzan 9\\
		Disk & 2MASS-GC01, BH 184 (Lynga 7), BH 261 (ESO 456-78), E 3 (ESO 37-1), FSR 1716, Mercer 5, NGC 104 (47 Tuc), & 24\\
		& NGC 4372, NGC 5927, NGC 6218 (M12), NGC 6256, NGC 6304, NGC 6352, NGC 6356, NGC 6362, NGC 6366, \\
		& NGC 6397, NGC 6441, NGC 6496, NGC 6539, NGC 6553, NGC 6569, NGC 6656 (M22), NGC 6749, NGC 6752, \\
		& NGC 6760, NGC 6838 (M71), NGC 7078 (M15), Pal 10, Pal 11, Pal 7 (IC 1276), Pal 8, Terzan 12, Terzan 3\\
		Sgr & Arp 2, NGC 6715 (M54)$^{*}$, Pal 12, Ter 7, Ter 8, Whiting 1 & 6\\
		Seq & FSR 1758, NGC 3201, NGC 5139($\omega$ Cen)$^{*}$, NGC 6101, NGC 6535 & 5\\
		Hel &  NGC 5024 (M53), NGC 5053, NGC 5272 (M3), NGC 5897, Pal 5 & 5\\
		GSE & IC 1257, NGC 362, NGC 1261, NGC 1851, NGC 1904 (M79), NGC 2298, NGC 4147, NGC 5286,  & 14\\
		& NGC 5694, NGC 6779 (M56), NGC 6981 (M72), NGC 7006, NGC 7089 (M2), Pal 2\\
		GSE-a & 2MASS-GC02, Djorg 1, NGC 2808, NGC 4833, NGV 6121 (M4), NGC 6554, UKS 1 & 7\\
		GSE-b & NGC 288, NGC 6205 (M13), NGC 7099 (M30), IC 4499 & 4\\
		GSE-c & ESO280-SC06, NGC 5634, NGC 6333 (M9), NGC 6584, NGC 6864 (M75)$^{*}$, Rup 106, Terzan 10 & 7\\
		Kraken & FSR 1735, NGC 5946, NGC 5986, NGC 6139, NGC 6254 (M10), NGC 6273 (M19)$^{*}$, NGC 6287, & 15\\
		& NGC 6402 (M14), NGC 6681 (M70), NGC 6541, NGC 6712, NGC 6809 (M55), Pal 6, Ton 2 (Pismis 26)\\
		Pot & AM 1 (E 1), AM 4, BH 140, Crater (Lae 1), Eridanus, Laevens 3, NGC 2419, NGC 4590 (M68), NGC 5466, & 26\\
		& NGC 5824, NGC 5904 (M5), NGC 6229, NGC 6235, NGC 6341 (M92), NGC 6426, NGC 6934, NGC 7492, Pal 1, \\
		& Pal 13, Pal 14 (Arp 1), Pal 15, Pal 2, Pal 3, Pal 4, Pyxis, VVV CL001\\
		\hline
    \multicolumn{2}{l}{Note. We will describe these progenitors in more detail in Section~\ref{sec:classificationofgcs}.}\\
    \multicolumn{2}{l}{$^{*}$The progenitors are nuclear star clusters (NSCs) of the old galaxies \citep[][]{pfeffer2020}.}\\
    \hline
	\end{tabular}
\end{table*}

\subsection{Method of classification}
\label{sec:method of classification}

\subsubsection{Age-metallicity relation}
\label{sec:age-metallicity relation}

We follow \citet{Massari2019} and \citet{Duncan2020} in classifying GCs through an age-metallicity relation (hereafter AMR, or AMRs for plural). A recent sample of 96 GCs with ages and metallicities is given by \cite{kruijssen2018} \footnote{The metallicity values presented in their paper are different from those presented in Section~\ref{sec:data} and Table~\ref{tab:A1}.}. For the final sample, the typical uncertainties are $\pm 0.75$\,Gyr for the age and $\pm 0.1$\,dex for the abundance of [Fe/H].
Results from these studies are affected by reddening, distance and chemical abundance \citep{VandenBerg_2013}, and due to the different data used to fit the isochrone lines, the age values in different studies cannot be simply combined. However, it is useful to get some values  from other studies \citep{bello2016, Oliveira_2020, Cohen_2021, Fern_ndez_2021, colmenares2021} for those GCs which are not included in the \citet{kruijssen2018} list.

\begin{figure}
	\includegraphics[width= \columnwidth]{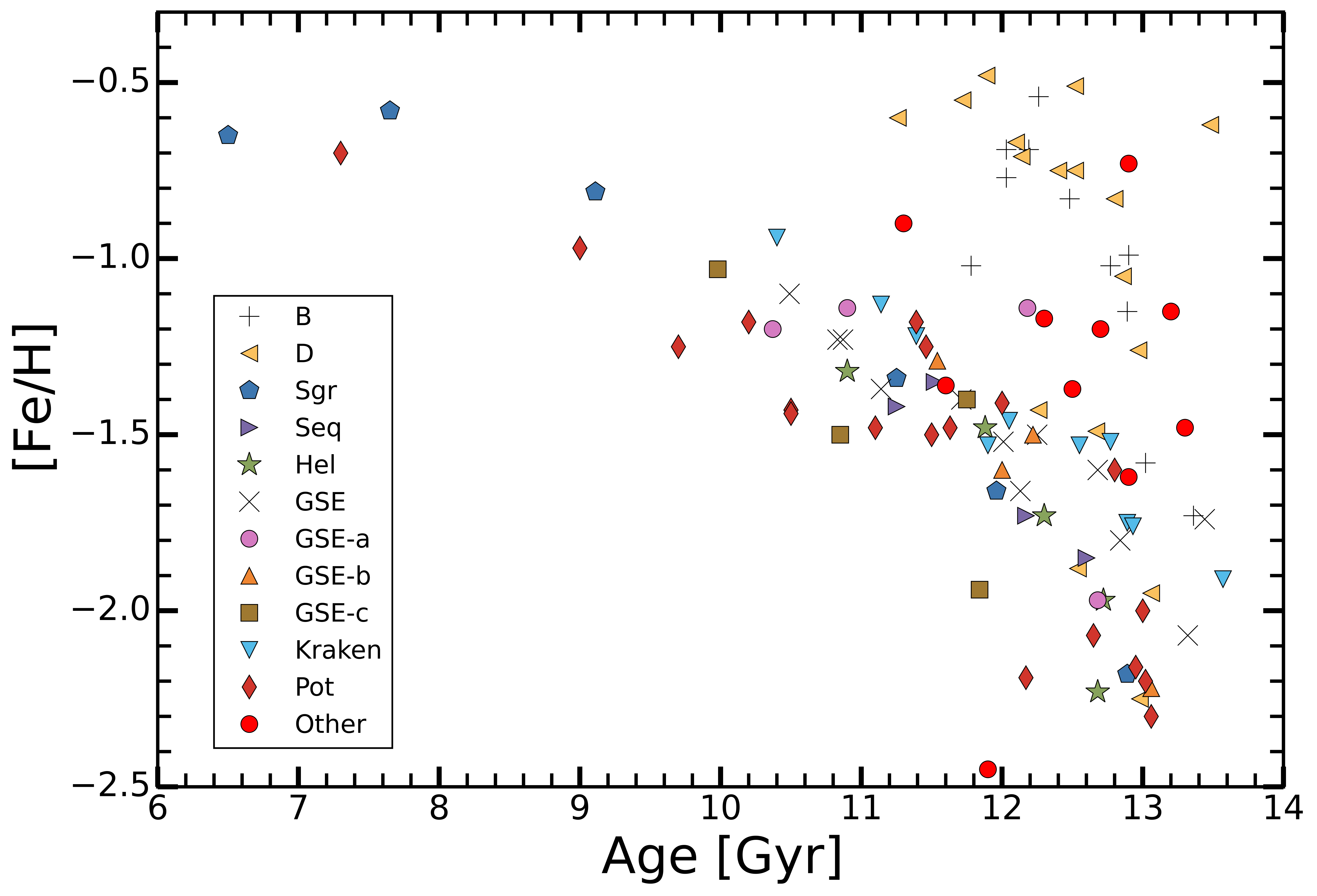}
    \caption{AMRs for the sample of 96 GCs, color-coded according to their associations with different progenitors. The black plus signs represent the Bulge GCs, the yellow leftward-pointing triangles represent Thick Disk GCs, the dark blue pentagons are associated with Sgr, the purple rightward-pointing triangles are Seq GCs, the green stars show Hel GCs, the black crosses indicate GSE GCs, the pink circle shows GSE-a GCs, the orange upward-pointing triangles signify GSE-b, the brown squares mark GSE-c, the sky blue downward-pointing triangles show Kraken GCs the red diamonds are Pot GCs. The red filled circles represent the age values from other studies.  
}
    \label{fig:fehago}
\end{figure}

The AMR is helpful in distinguishing in situ GCs from ex situ GCs. We plot the AMR for 96 GCs in Fig.~\ref{fig:fehago}. Most age values we used are from \cite{kruijssen2018}.
Different groups are labeled with different colors and shapes. This figure shows that the GCs fall into two apparent types.
One is the in situ GCs (bulge and thick disk GCs,\footnote{Since the thin disk does not contain GCs, we refer to thick disks as just disks hereafter.} listed as B and D respectively in the figure),  which is located on the young and more metal-rich branch of the AMR. The other is the accreted GCs, those located on the young and metal-poor branch.  Since both accreted and in situ GCs overlap in the region of the AMR space, the origin of the oldest GCs cannot be distinguished based on the AMR alone.

The AMR can be understood with a leaky-box chemical evolution model \citep{Prantzos2008, kruijssen2018, Massari2019}, which reflects the evolution of metallicity over time in primordial galaxies:
\begin{equation}
    [\mathrm{Fe/H}] = -p\ln{\frac{t_f-t}{t_f-t_i}}
	\label{eq:amr}
\end{equation}
where $p$ is the effective yield. In general, a larger $p$ indicates a larger initial mass of the galaxy.  $t_f$ represents the age of the universe (where the time of the Big Bang is 0). This expression is obtained by assuming a constant star formation rate from the start time $t_i$ to the end time $t_f$. 

\begin{figure}
	\includegraphics[width= \columnwidth]{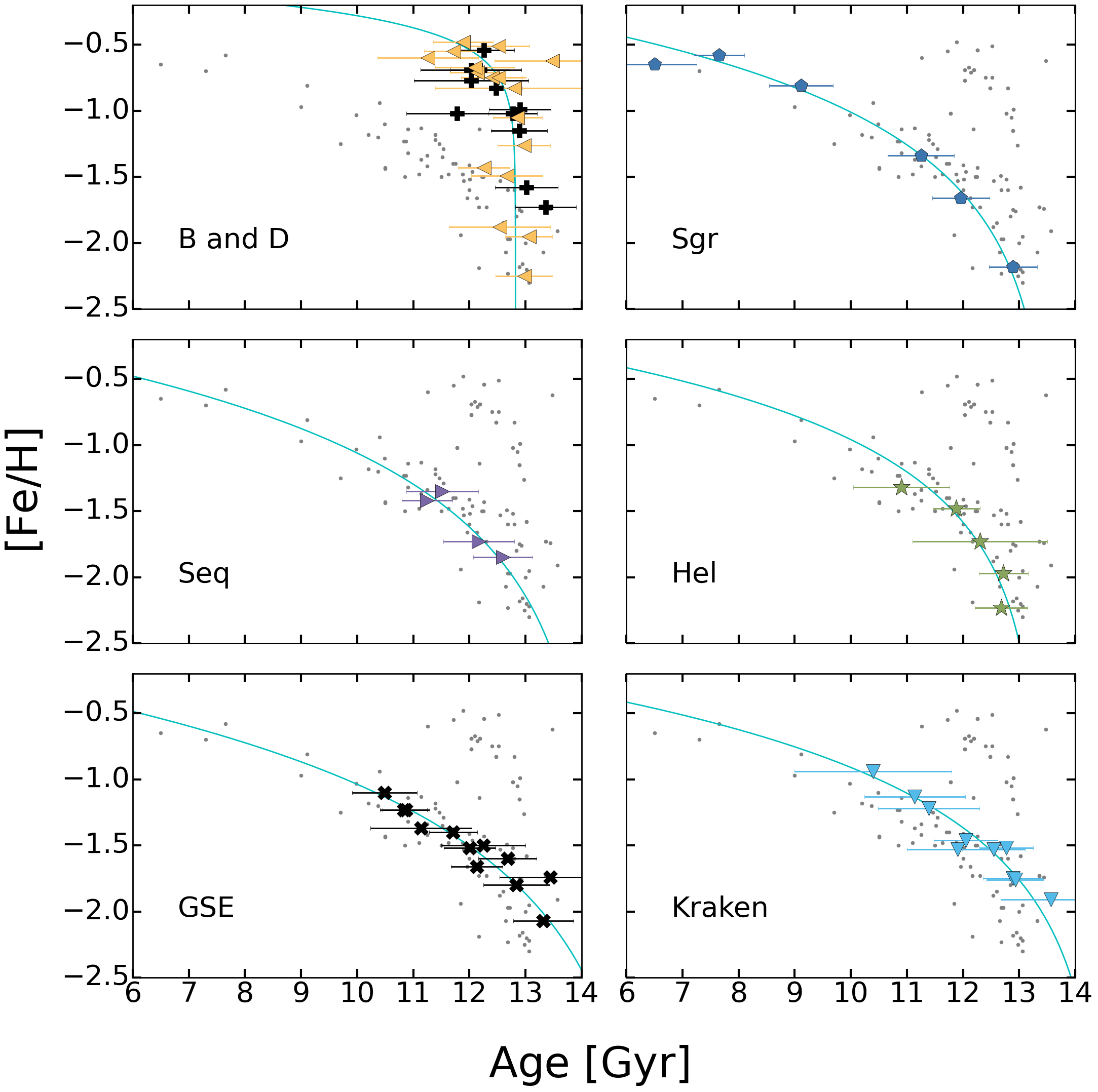}
    \caption{AMRs for individual progenitors with age estimates. Color-coding is the same as in Fig.~\ref{fig:fehago}.  In each panel, the corresponding progenitors are given in the lower left corner.  For easy viewing, bulge and GSE GCs are in bold. The gray dots represent other clusters that do not participate in the fit. In addition,  the age uncertainty of individual objects is also plotted. The resulting curve for each progenitor is signified by the solid blue line.}
    \label{fig:fehdu}
\end{figure}

In Fig.~\ref{fig:fehdu}, we display the metallicity distribution with the age. The relationship in Equation~\ref{eq:amr} for each progenitor is given by the blue solid curve. It is clear that a simple leaky box chemical evolution model can describe the AMRs well with the reasonably the effective yield and time. However, since the possible error is too large, we do not show the fitting results. This fit is only used to motivate our classification. The AMR is not the only or even the main method we employ for classifying the different GCs  because some GCs mix together in the AMR diagram of Fig.~\ref{fig:fehdu}.



\subsubsection{Orbits and Integrals of motion}
\label{sec:integrals of motion}

The AMRs for most ex situ GCs are intertwined, with even some in situ and ex situ GCs overlapping in areas of low metallicity. It is difficult to find the progenitors of GCs by relying only on AMRs. Of course, this problem also exists in other methods described in this section. It is necessary, therefore, to combine different methods to make more accurate judgements on the progenitors of GCs.  As proposed by \citet{Bajkova2020}, GCs can be divided into different subsystems based on the bimodal distribution of GCs over the parameter $L_z/\mathrm{ecc}$.

\begin{figure*}
	\includegraphics[width= \textwidth]{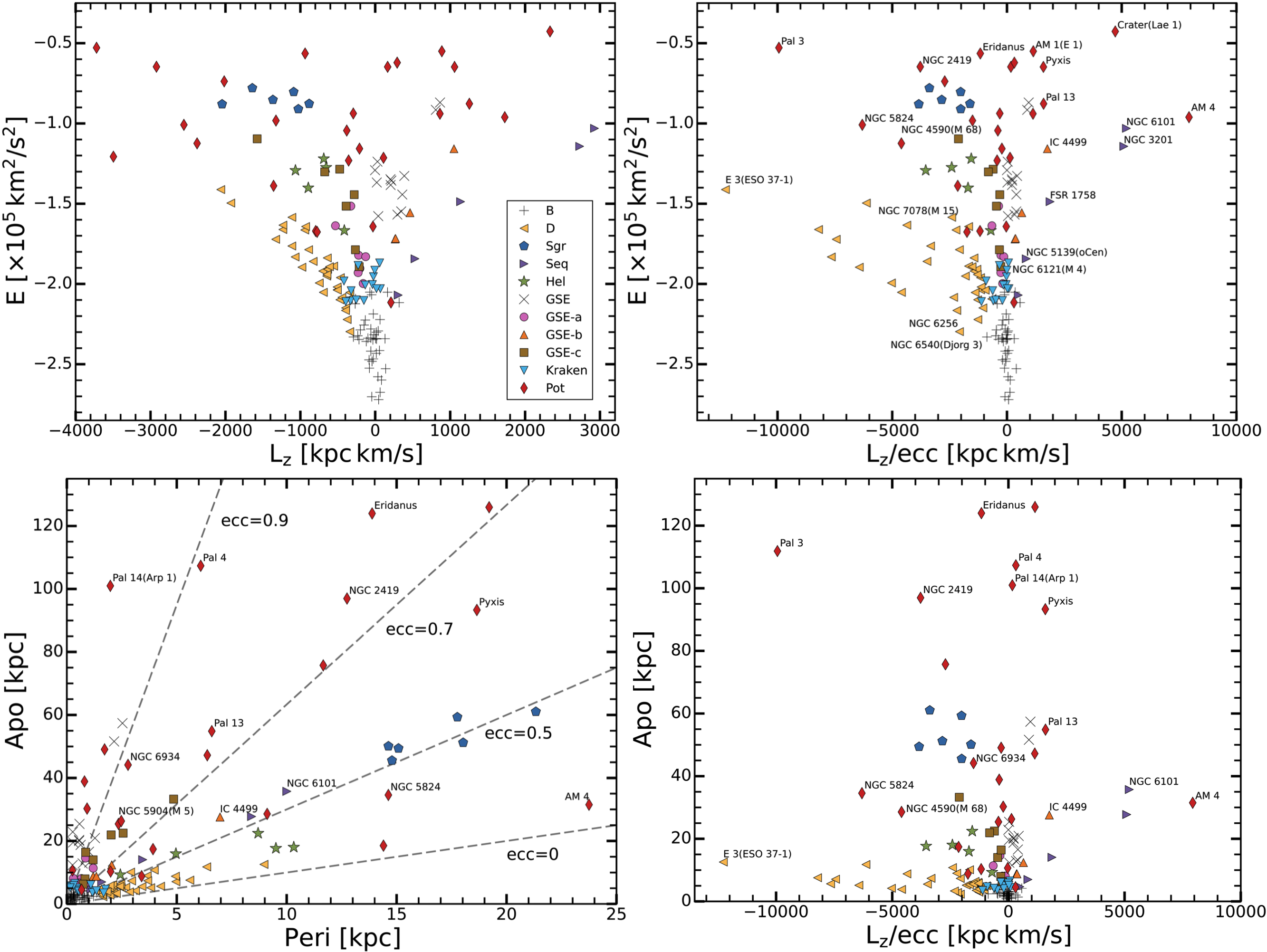}
    \caption{Two-dimensional diagrams $L_z -E$ (upper left), $L_z/\mathrm{ecc}-E$ (upper right), Peri-Apo (lower left), $L_z/\mathrm{ecc}$-Apo (lower right) for the 159 GCs in our sample. Colour-coding is the same as in Fig.~\ref{fig:fehago}.
    We have labelled some interesting GCs.
    For visualization purposes, clusters Pal 1 with extremely negative $L_z/\mathrm{ecc}$, and Pal 3 and Crater with extremely positive pericentres have not been plotted.}
    \label{fig:zongtu}
\end{figure*}

For each GC, there are errors in the integrals of motion (hereafter IOM) $E$, $L_z$, ecc, Peri and Apo from the measurement errors in its position and velocity. In order to estimate these errors, for each GC we generate a sample of 2000 perturbed initial conditions by adding to its position and velocity a random noise following a Gaussian distribution with the specified variance, then integrating its orbit with the initial condition for 20 Gyr or 50 cycles, whichever is smaller.   
The errors for $E$, $L_z$, ecc, Peri and Apo are given by the standard deviation of these 2000 orbits for each GC. 

In Fig.~\ref{fig:zongtu}, we feature the two-dimensional diagrams for $L_z-E$ (upper left), $L_z/\mathrm{ecc}-E$ (upper right), Peri-Apo (lower left) and $L_z/\mathrm{ecc-Apo}$ (lower right).  It is clear that the GCs can be separated well in the $L_z-E$ and $L_z/\mathrm{ecc}-E$ diagrams. The bulge GCs are located in the central region of the MW and their orbits deviate significantly from circular since the bar/bulge dominates the potential in this central region of the MW, so the bulge GCs have small $L_z$ and low energy. For the disk GCs, they have nearly circular orbits and larger $L_z$ than the bulge GCs. Compared with the $L_z-E$ diagram, the $L_z/\mathrm{ecc}-E$ diagram gives a more extended distribution for the disk GCs because the orbits of the disk GCs are nearly circular and their eccentricities are small. The disk GC distribution is also clear in the diagram of $L_z/\mathrm{ecc-Apo}$. The Peri-Apo diagram shows the ecc distribution of GCs with different progenitors, which again indicates that the disk GCs have relatively small ecc values.

Most of the data in Fig.~\ref{fig:zongtu} have slight errors. Only the higher energy data have significantly large errors. It is even hard to determine the direction of their rotation (positive or negative of $L_z$). In addition, since the parameters in Fig.~\ref{fig:zongtu} depend on the MW model, we also calculate the GC orbits by using five other models, which are abbreviated as Pi14 \citep{Piffl2014}, Bo15 \citep[][\texttt{MWPotential2014}]{Bovy2015}, Bi17 \citep{Binney2017}, and Pr19\footnote{MilkyWayPotential from Gala \citep{Price-Whelan2017}, whose disk model is taken from Bo15.}. We find that different MW gravitational potential models do not have a significant effect on the classification of the GCs.


\begin{figure}
	\includegraphics[width= \columnwidth]{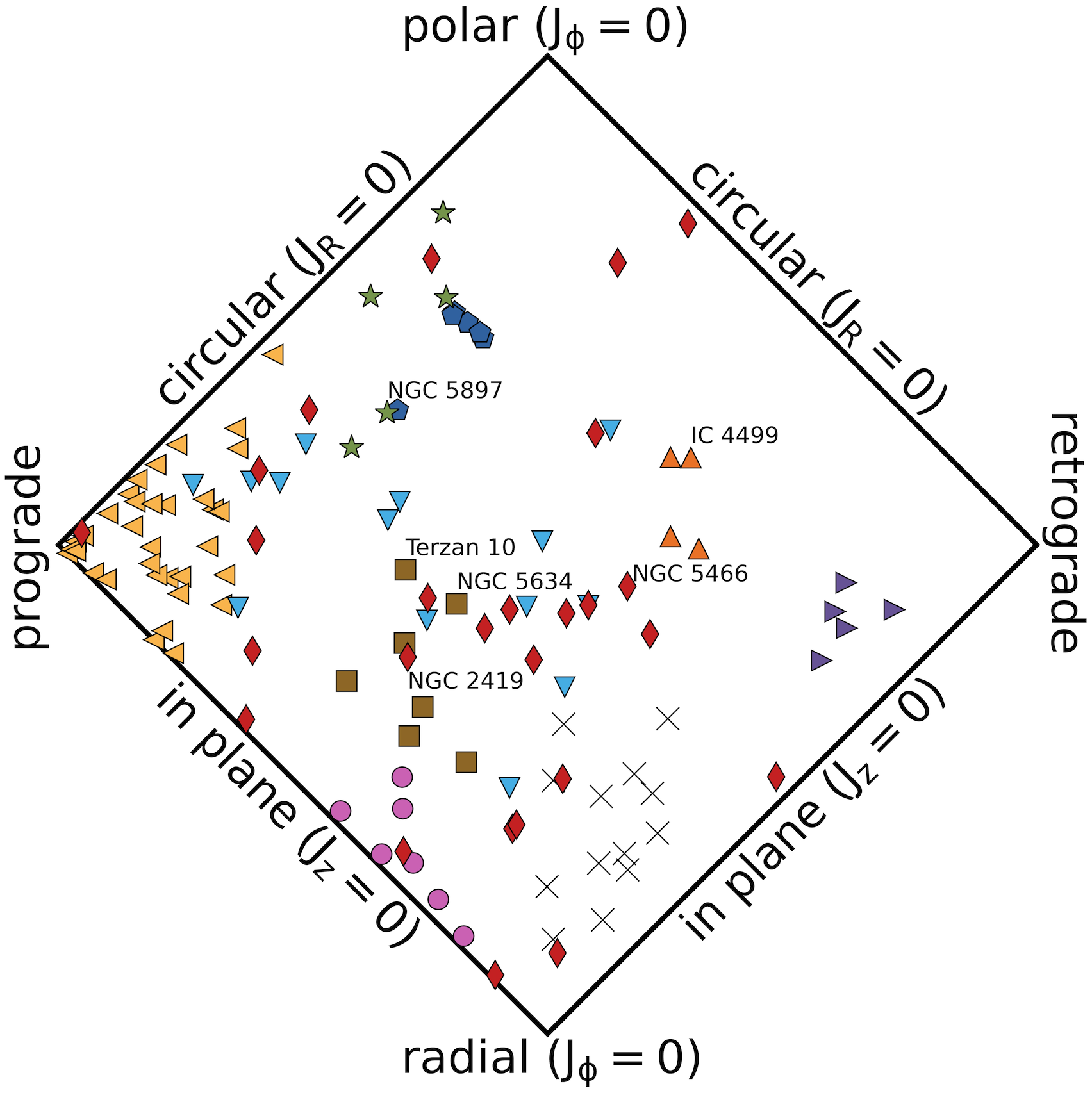}
    \caption{The action space map for 121 GCs. Color-coding is the same as in Fig~\ref{fig:fehago}. The figure is similar to Fig.5 in \citet{myeong2019} and Fig.5 in \citep{Vasiliev2019}. The horizontal axis is $J_\phi/J_{\rm tot}$, and its range is from -1 (in the left-hand corner) to 1 (in the right-hand corner). The vertical axis is $(J_z-J_R)/J_{\rm tot}$, and its range is from -1 (bottom) to 1 (up).}
    \label{fig:space}
\end{figure}

Fig.~\ref{fig:space} displays the action space map for 121 MW GCs, excluding those in the bulge region. The bulge GCs are distributed evenly throughout the map. The horizontal axis is $J_\phi/J_{\rm tot}$, while the vertical axis is $(J_z-J_R)/J_{\rm tot}$, where $J_{\rm tot} = \lvert J_R \rvert + \lvert J_\phi \rvert + \lvert J_z \rvert$ is the sum of absolute values of all three actions.  This plot can be viewed as a projection of energy onto the $J_R$, $J_\phi\,(\equiv L_z)$, $J_z$ spaces \citep[][]{binney2011galactic}. Different positions of points in the graph represent different weights of action. The GCs with retrograde orbits are on the far left of the diagram, while GCs with prograde orbits are on the far right.

The action space map can help us to identify the Disk, Sgr, Seq and GSE GCs. In addition, they are also helpful for the subsequent GSE separation. However, the Kraken and the Hel GCs are scattered in the action space map, and cannot be classified in this way.  

\begin{figure*}
	\includegraphics[width= \textwidth]{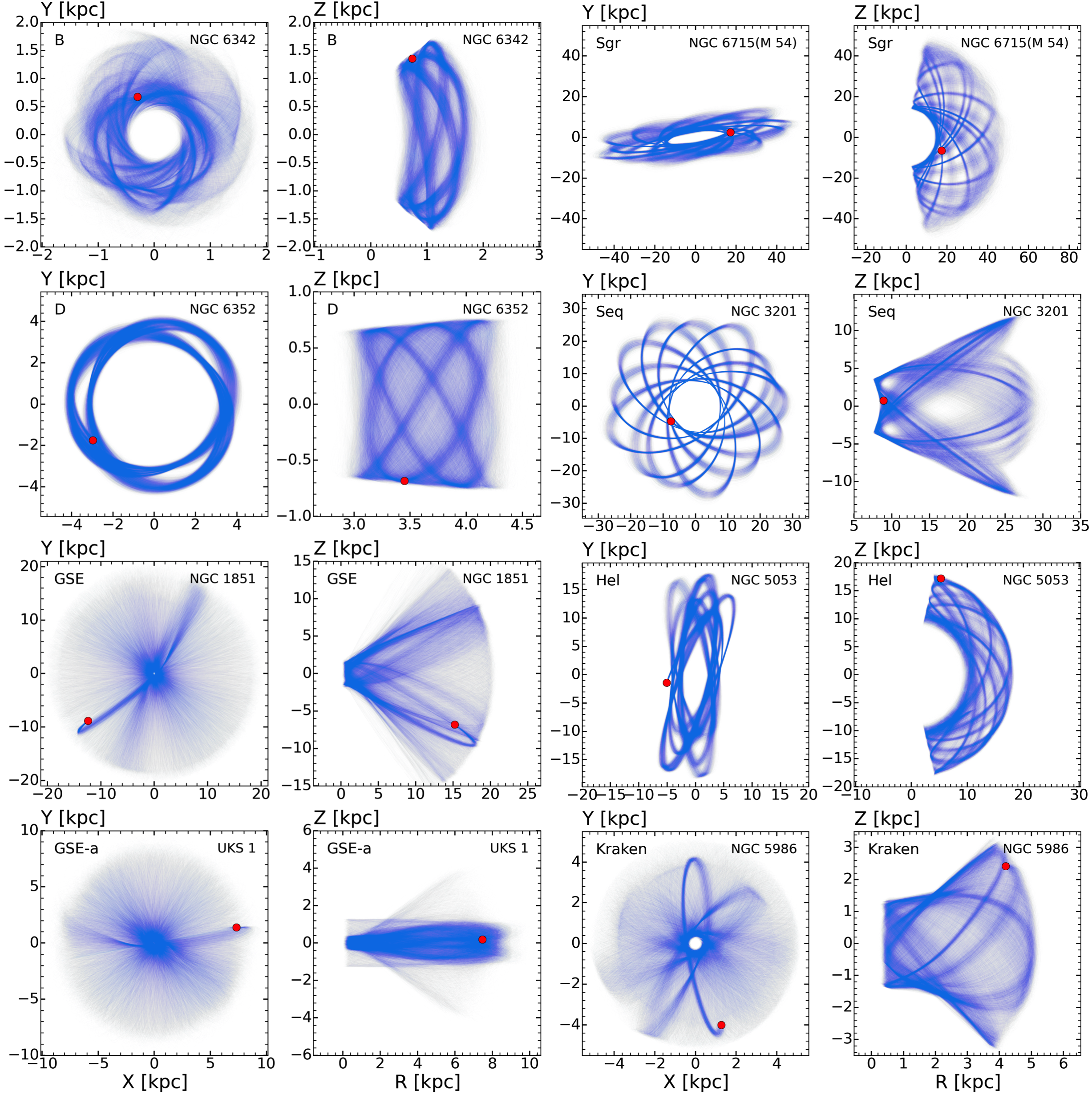}
    \caption{Orbits of typical GCs in different families. The plot in each panel is the superposition of 2000 slim orbits, and their initial conditions are randomly generated by considering the errors of the position and velocity. To give a clear orbit structure for each GC, each orbit is shown for only 10 orbital periods. The initial position of the orbit is marked with a red circle. In each panel, the upper left is marked with the progenitors of the GC, and the upper right is the name of the GC. For each GC, the orbits are drawn in both the $X$-$Y$ and $R$-$Z$ planes.}
    \label{fig:orbit}
\end{figure*}

Fig.~\ref{fig:orbit} exhibits some typical examples of GC orbits. Each GC orbit is shown on both the $X$-$Y$ and $Z$-$R$ planes. For each GC,  we randomly generate 2000 initial conditions and integrate orbits for all initial conditions. Therefore, the orbits displayed in this plot for each GC are the superposition of 2000 orbits and each orbit is drawn with a thin line. The blue regions have relatively high probability, while the gray parts are the relatively low probability regions. It is seen that the GCs with the same progenitors have similar orbit morphology. Bulge GCs, such as NGC 6638, have a relatively small radius $R$. Disk GCs, such as NGC 6352, have a small $z$ range. For more examples, the $R$-$Z$ plots of Sgr and Hel show large $z_{\rm max}$ values. The $X$-$Y$ plots of the GSE and GSE-a clusters suggest large radial velocities and higher orbital eccentricities.

However, it neither means that all clusters in a category have a particular orbital shape, nor that GCs with this orbital shape necessarily belong to the progenitor. 

\subsection{Membership}
\label{sec:classificationofgcs}
We classify the GCs by taking into account of information from multiple methods. The quantitative criteria we set are based on an evaluation of all presently available data, which we think can best distinguish the different GCs, and should not be regarded as  inflexible rules. In some cases, conflicting or ambiguous results are obtained from criteria based on different methods, so the GC meeting a specific criterion of a progenitor is not always assigned to this progenitor, as we have to also consider the fitting to criteria of other methods. In the following, we discuss the reference criteria for the different sources and the resulting membership for the  GCs based on the methods described above.

\begin{table*}
	\centering
	\renewcommand{\arraystretch}{1}
    \setlength\tabcolsep{5pt}
	\caption{Different classifications of GCs between our results and those in \citet[hereafter M19]{Massari2019}. The result of \citet[hereafter BB21]{2021RAA....21..173B} and \citet[hereafter C22]{2022arXiv}, are also given for the reference. For the convenience of viewing, we use the abbreviation of the progenitor here.}
	\label{tab:duibi}
	\begin{tabular}[c]{lcccc|lcccc}
		\hline
		Name & Progenitor & M19 & BB21 & C22 & Name & Progenitor & M19 & BB21 & C22\\
        \hline

BH 261 (ESO 456-78) & D      & B       & D     & Kraken & NGC 6341 (M 92)    & Pot   & GSE   & GSE   & GSE    \\
E 3 (ESO 37-1)      & D      & Hel     & D     & D      & NGC 6401          & B     & L-E   & Seq   & B      \\
IC 4499            & GSE-b  & Seq     & Seq   & Seq    & NGC 6441          & D     & L-E   & L-E   & D      \\
NGC 2419           & Pot    & Sag     & Sag   & Pot    & NGC 6453          & B     & L-E   & L-E   & B      \\
NGC 4590(M 68)     & Pot    & Hel     & Hel   & Hel    & NGC 6517          & B     & L-E   & L-E   & B      \\
NGC 5466           & Pot    & Seq     & GSE   & Seq    & NGC 6539          & D     & B     & D     & Kraken \\
NGC 5694           & GSE    & H-E     & GSE   & Seq    & NGC 6544          & GSE-a & L-E   & GSE   & Kraken \\
NGC 5824           & Pot    & Sag     & Hel   & Pot    & NGC 6553          & D     & B     & D     & Kraken \\
NGC 5897           & Hel    & GSE     & GSE   & GSE    & NGC 6569          & D     & B     & D     & Kraken \\
NGC 5904 (M 5)      & Pot    & Hel/G-E & GSE   & Hel    & NGC 6584          & GSE-c & H-E   & GSE   & Hel    \\
NGC 6093 (M 80)     & B      & L-E     & L-E   & B      & NGC 6981 (M 72)    & GSE   & Hel   & GSE   & Hel    \\
NGC 6121 (M 4)      & GSE-a  & L-E     & L-E   & Kraken & NGC 7006          & GSE   & Seq   & GSE   & Seq    \\
NGC 6144           & B      & L-E     & Seq   & B      & NGC 7492          & Pot   & GSE   & GSE   & Hel    \\
NGC 6229           & Pot    & GSE     & GSE   & Hel    & Pal 1             & Pot   & D     & D     & Pot    \\
NGC 6235           & Pot    & GSE     & D     & GSE    & Pal 2             & Pot   & GSE   & GSE   & Pot    \\
NGC 6256           & D      & L-E     & L-E   & B      & Pal 6             & B     & L-E   & L-E   & B      \\
NGC 6284           & Kraken & GSE     & GSE   & Kraken & Pal 13            & Pot   & Seq   & GSE   & Seq    \\
NGC 6304           & D      & B       & D     & B      & Pal 15            & Pot   & GSE   & GSE   & Seq    \\
NGC 6333 (M 9)      & GSE-c  & L-E     & L-E   & Kraken & Rup 106           & GSE-c & Hel   & Hel   & Hel    \\

		\hline
	\end{tabular}
\end{table*}

\subsubsection{Bulge}
\label{sec:bulge}

In Figs.~\ref{fig:fehago}--\ref{fig:space}, the bulge clusters are labeled by the black plus signs. 
For the bulge GCs, since they are located in the central region of the MW, they should have small Apo, low energy and low angular momentum. Therefore, we simultaneously use Apo, $E$ and $L_z$ to select the bulge GCs. These GCs should satisfy
\begin{equation}
\begin{aligned}
    (&~\mathrm{Apo}<4~\rm{kpc})\\
    \land  &~ (E<-1.9\times10^5~\rm{km^2\,s^{-2}})\\
    \land &~ (-1000<L_z/\mathrm{ecc}<1000~\rm{kpc\,km\,s^{-1}})
\end{aligned}
\label{eq:bugle}
\end{equation}
where $\land$ means the logical sum. Here we take a different critical Apo value from that used in M19 (Apo\,$<3.5$\,kpc). There are two reasons: 
First, the MW potential we adopted here is different from that in M19 and the Apo of the orbits depends on the potential. Second, we can reassess the best Apo value by taking into account the information provided by other methods, e.g. the AMR, which can also help distinguish the bulge clusters from others. In Fig.~\ref{fig:zongtu}, we find that some Kraken GCs have Apo, $E$ and $L_z$ similar  to those of the bulge GCs, making it difficult to distinguish the bulge GCs and Kraken GCs through Apo, $E$ and $L_z$, but we can distinguish part of them by AMR. In Fig.~\ref{fig:fehago} and Fig. ~\ref{fig:fehdu}, we can see that bulge GCs and Kraken GCs have obvious difference in their AMRs.   Noting that NGC 6093 (M80), NGC 6144, NGC 6171 (M107), NGC 6388, NGC 6723 satisfy the in situ AMR and have relatively large Apo in our MW model, we increase the critical Apo from 3.5 kpc to 4 kpc.  For GCs without age information, we only use Apo, $E$ and $L_z$ to determine whether they belong to the bulge GCs.   

Recently, \citet{ferraro2021} found a GC-like system in the bulge region, named Liller 1.  This system has two distinct stellar populations with significantly different ages, 1-3 Gyr for the youngest and 12 Gyr for the oldest. The oldest stellar population in GC Liller 1  is similar to the old component of Terzan 5 (11), which is also a GC system hosting at least two major subpopulations \citep[][]{ferraro2009, Lanzoni_2010, Massari_2014, Ferraro_2016}. Combining all available observational data, \citet{ferraro2021} suggested that both Liller 1 and Terzan 5 (11) are possible remnants of primordial in situ clumps, originating from the fragmentation of the early giant structures, that sunk toward the central region of the MW because of dynamical friction and built-up the bulge through coalescence.

There are also some other interpretations that demonstrate the complexity of Terzan 5 (11), such as a possible collision between the metal-poor Terzan 5 (11) and a giant molecular cloud \citep{mckenzie2018}, which would have resulted in a new generation of stars. \citet{bastian2022} also preferred this scenario and predicted that these GCs should have high masses, and have disk-like orbits in the inner regions of the MW, which are consistent with what is seen in Terzan 5 (11) and Liller 1.

\subsubsection{Disk}
\label{sec:disk}

Shown as yellow leftward-pointing triangles in Figs.~\ref{fig:fehago}--\ref{fig:space}, the disk GCs are defined as 
\begin{equation}
\begin{aligned}
   & ~ (E<-1.5\times10^5~\rm{km^2\,s^{-2}})\\
    &\land ~ (L_z/\mathrm{ecc}<-800~\rm{kpc\,km\,s^{-1}})\\
    &\land ~ (J_\phi/J_{\rm tot}<-0.5)\\
    &\land ~ (z_{\rm max}<5~\rm{kpc})\\
    &\land ~ (\mathrm{ecc} <0.5)
\end{aligned}
\label{eq:disk}
\end{equation}
Six bulge GCs (BH 261 (ESO 456-78), NGC 6304, NGC 6539, NGC 6553, and NGC 6569) and a Kraken (L-E) GC (NGC 6256) in M19 are considered more likely to be disk GCs because of their extremely high $\lvert L_z \rvert$. These six GCs are also recorded as disk GCs in BB21 (see Table~\ref{tab:duibi}).
In addition, we note that E 3 (ESO 37-1) and Pal 10 have high energies and it is difficult to distinguish them from Hel clusters in the $L_z-E$ diagram. However, in the $L_z/\mathrm{ecc}-E$ distribution, these two GCs are significantly separated from Hel GCs and closer to disk GCs. In particular, the AMR of GC E 3 (ESO 37-1) fits well with in situ clusters, so our disk directory includes E 3 (ESO 37-1) and Pal 10.

The ecc values of GCs NGC 6656 (M22), NGC 6749 and Pal 8 are greater than our ecc threshold value. Considering the IOM space, we assess these three GCs as disk GCs. The classifications for the former two are consistent with those in M19. NGC 6656 (M22) has the highest ecc (ecc=0.56) in our disk GCs.   
Compared with the Disk GCs in M19, we also include two new clusters Mercer 5 and 2MASS-GC01.

\subsubsection{Sagittarius dwarf (Sgr)}
\label{sec:sag}

Shown as dark blue pentagons in Figs.~\ref{fig:fehago}--\ref{fig:space}, the Sgr dwarf galaxy is the first known MW accretion object \citep[][]{ibata1994}. \citet{Law_2010a} developed numerical models of Sgr's tidal disruption, and they found that the initial Sgr system may have included 5-9 GCs.
A total of 23 GCs may have been associated with the Sgr, where five (Arp 2, NGC 6715 (M54), NGC 5634, Terzan 8 and Whiting 1) have high probability, four (Berkeley 29, NGC 5053, Pal 12, and Terzan 7) have medium probability and two (Pal 12, and Terzan 7) have low probability \citep{Law_2010a}.           
Many subsequent studies \citep[][]{massari2017, Sohn2018, bellazzini2020} have improved the classification of these GCs based on new observations such as Gaia DR2 and Hubble Space Telescope (HST). Recently, the discovery of multiple low luminosity GC candidates  \citep[][]{minniti2021, minniti2021a} possibly associated with Sgr may help us to improve understanding of Sgr.

The GCs associated to Sgr are given by 
\begin{equation}
\begin{aligned}
    & ~ (-1.0\times10^5<E<-0.6\times10^5~\rm{km^2\,s^{-2}})\\
    &\land ~ (-3000<L_z<-500~\rm{kpc\,km\,s^{-1}})\\
    &\land ~ (3500<L_\perp<6000~\rm{kpc\,km\,s^{-1}})\\
    &\land ~ (-0.5<J_\phi/J_{\rm tot}<0)\\
    &\land ~ (0.2<(J_z-J_R)/J_{\rm tot}<0.6)\\
\end{aligned}
\label{eq:sag}
\end{equation}
where $L_\perp=\sqrt{J_R^2+J_z^2}$ is the angular momentum component perpendicular to $L_z$. 

For NGC 2419, \citet{Sohn2018} and \citet{2021RAA....21..173B} took it as part of Sgr. However, we classified it into the Pot group, because its IOM and AMR are not consistent with the characteristics of Sgr GCs, but more closely aligned with the Pot group.         
From Fig.~\ref{fig:space}, we can see that most Sgr GCs are concentrated in the action space, and the only significant deviation from the concentrated region is from Pal 12. This GC satisfies our selected conditions for Sgr GCs, and is also a Sgr GC according to M19.

\subsubsection{Sequoia galaxy (Seq)}
\label{sec:seq}

Shown as purple rightward-pointing triangles  in Figs.~\ref{fig:fehago}--\ref{fig:space},  
the Seq GCs are named by their progenitor dwarf galaxy Sequoia, which provides many high energy retrograde stars in the stellar halo and may include six GCs \citep[][]{2018MNRAS.475.1537M, myeong2018, 2018ApJ...856L..26M, 2018ApJ...863L..28M, myeong2019}. Recently, \citet{feuillet2021} went into more detail about the motion and chemical limitations of the stars in Seq, and \citet{matsuno2021} measured the abundance of various elements in many stars using high-resolution spectroscopy. Based on these data, the differences between Seq, GSE and in situ stars were determined.
If more GC chemical abundance data can be acquired in the future, the distinction among them should become more clear.

The most notable feature of the Seq GCs is their retrograde orbits.  Moreover, the Seq GCs have a typical orbit ellipticity of   $\mathrm{ecc} \sim 0.6$. The Seq GCs are selected as follows,
\begin{equation}
\begin{aligned}
    &~ (L_z>0~\rm{kpc\,km\,s^{-1}})\\
    &\land ~ (J_\phi/J_{\rm tot}>0.3)\\
    &\land ~ (-0.4<(J_z-J_R)/J_{\rm tot}<0.1)\\
    &\land ~ (\mathrm{ecc} \sim 0.6)
\end{aligned}
\label{eq:seq}
\end{equation}
We have not given an energy range for selecting the Seq GCs because the energy span for this kind of GC is relatively wide. The selected GCs are well determined by the other conditions, and there is no need to use energy to give further constraints. 

We assess four possible Seq GCs in M19 (NGC 3201, NGC 5139 ($\omega$Cen), NGC 6101, NGC 6535) according to the above criteria. These clusters were given two possible progenitors in M19. In Fig.~\ref{fig:space}, IC 4499 deviates significantly from the other possible Seq clusters. As we will show in Section~\ref{sec:Two}, we consider IC 4499 to be consistent with the GSE-b criterion. As NGC 5466 and Pal 13 deviate from other Seq GCs in both IOM and action images, and cannot be associated with other known accretion structures, we put them in the Pot family.

\subsubsection{Helmi Streams (Hel)}
\label{sec:hel}

Shown as green stars in Figs.~\ref{fig:fehago}--\ref{fig:space}, this progenitor was proposed by \cite{HELMI1999} as a kinematically coherent group of stars. Recently, \citet{koppelman2019} carried out a detailed study of Hel, and obtained seven related GCs. The Hel GCs are defined as 
\begin{equation}
\begin{aligned}
    & ~ (-1.6\times10^5<E<-1.0\times10^5~\rm{km^2\,s^{-2}})\\
    &\land ~ (-1500<L_z<-300~\rm{kpc\,km\,s^{-1}})\\
    &\land ~ (1000<L_\perp<3200~\rm{kpc\,km\,s^{-1}})\\
    &\land ~ (-0.7<J_\phi/J_{\rm tot}<0)\\
    &\land ~ ((J_z-J_R)/J_{\rm tot}>0)
\end{aligned}
\label{eq:hel}
\end{equation}
Based on our criteria, and taking into account the results from other studies (M19, BB21, and C22), NGC 5024 (M53), NGC 5053, NGC 5272 (M3) and Pal 5 are safe Hel GCs. In addition, as affirmed in Table~\ref{tab:duibi}, for other Hel GCs in M19, such as E 3 (ESO 37-1), NGC 6981 (M72), Rup 106, NGC 5634 and NGC 5904 (M5), we assigned them to other progenitors. Furthermore, since NGC 4590 is difficult to be associated with any known structure in this paper, it is classified as a Pot GC. Note that two GCs (Rup 106 and NGC 5634) are assigned to GSE-c, and, as we will show in Section~\ref{sec:Three}, there are some overlaps between Hel and GSE-c. The most obvious difference between the two is their ellipticities and their positions in action space. In the end, we only allocated five GCs for Hel.

\subsubsection{Gaia-Sausage-Enceladus}
\label{sec:ge}

Shown as black crosses in Figs.~\ref{fig:fehago}--\ref{fig:space}, Gaia-Enceladus, as proposed by \citet{helmi2018}, is a major merger event in the MW's history, which may directly lead to the formation of the MW's thick disk and halo. 
At the same time, \citet{belokurov2018} found a similar metal-poor and extreme radial orbital structure named Gaia-Sausage. After that, \citet{2018ApJ...863L..28M} described the kinematic characteristics of the Gaia-Sausage. These two events are often combined and called GSE. As we know, the GSE includes both prograde and retrograde GCs, which are explained as the result of a merger between a giant disk galaxy and the MW \citep{koppelman2020}. 
However, \citet{Kwang_Kim_2021} found evidence that the GSE is the result of multiple accretion events, which is also supported by our own results of the AMR and orbit properties of GCs. 

Considering the results of \citet{Kwang_Kim_2021} and the distinct characteristics of GSE GCs in the phase space, we divided the original GSE GCs into four parts. We have extracted three new structures (GSE-a, GSE-b and GSE-c) from the previously defined GSE, and the remaining main part is still referred to as GSE GCs. The GSE GCs are selected by
\begin{equation}
\begin{aligned}
    & (-1.7\times10^5<E<-0.8\times10^5~\rm{km^2\,s^{-2}})\\
    &\land ~ (-100<L_z<800~\rm{kpc\,km\,s^{-1}})\\
    &\land ~ (-0.1<J_\phi/J_{\rm tot}<0.4)\\
    &\land ~ ((J_z-J_R)/J_{\rm tot}<0.5)\\
    &\land ~ (\mathrm{ecc} > 0.8)
\end{aligned}
\label{eq:ge}
\end{equation}
Based on the IOM diagram in Figs.~\ref{fig:space}, we find that NGC 5694, NGC 6981 (M72), and NGC 7006 deviate significantly from their assigned progenitors in M19, and these clusters can be interpreted as GSE GCs. 
GSE retains those GCs with the strongest constraints, and the three new structures presented in this paper are those with weaker constraints in GSE phase space.

\begin{figure*}
	\includegraphics[width= \textwidth]{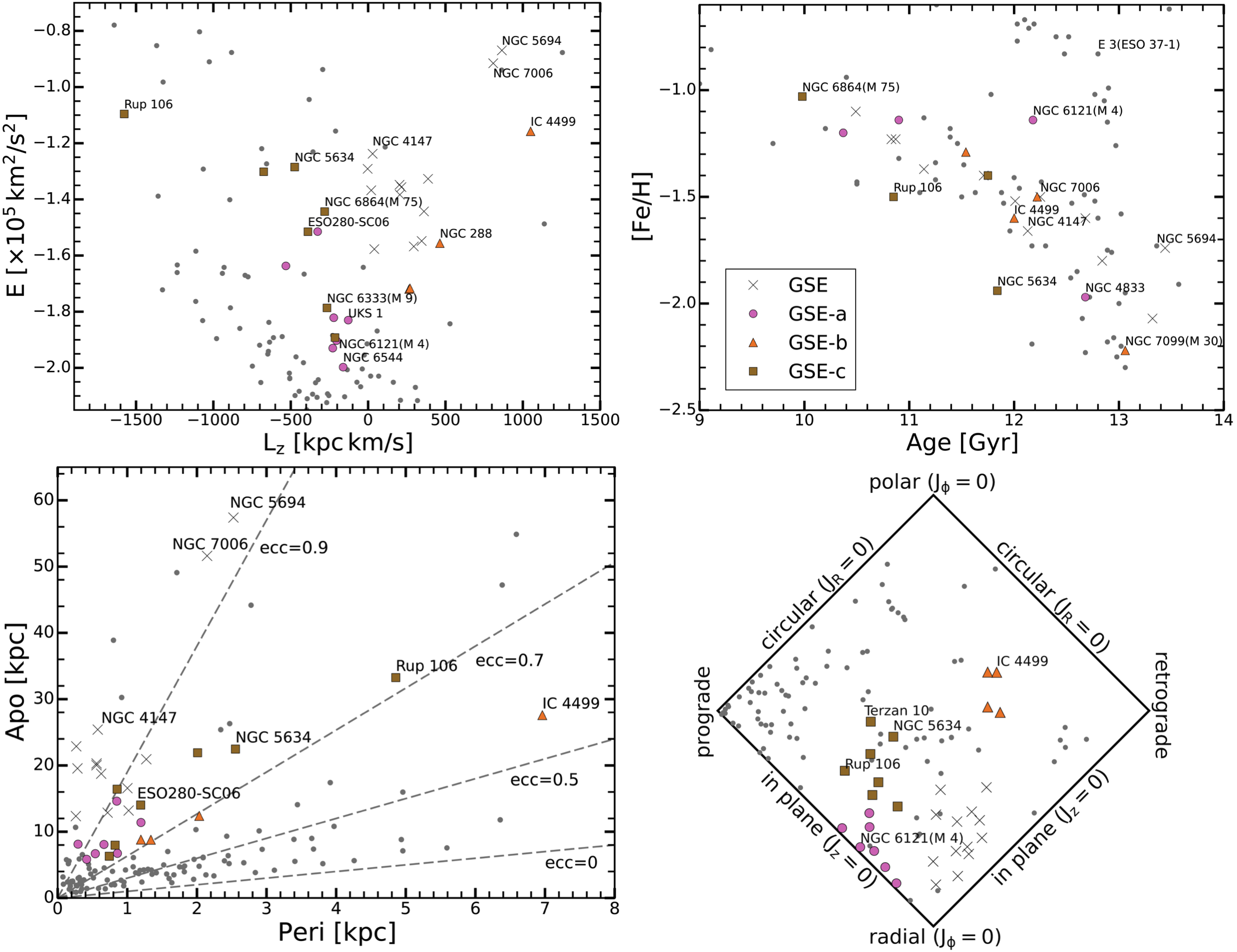}
    \caption{GCs of GSE and three new branches split from the GSE. Here we show four images that are crucial to distinguishing these clusters, which are $L_z$-E (upper left), Age-[Fe/H] (upper right), Peri-Apo (lower left) and action space (lower right).  Since NGC 6121 (M4) significantly deviates from GSE-a in the upper right panel, it is specially marked. In addition to that, these GCs are well separated into the four groups.}
    \label{fig:new}
\end{figure*}

\subsubsection{GSE-a}
\label{sec:one}


Shown as pink circles in Figs.~\ref{fig:fehago}--\ref{fig:space}, these are associated with the first possible accretion event we have isolated from GSE. The most striking feature of these GCs is their orbits which are close to the equatorial plane of the MW. These orbits have very low $z_{\rm max}$ values (about 1 kpc).
GSE-a exhibits characteristics different from typical GSE GCs. However, small $z_{\rm max}$ does not fully account for an accretion event in the history of the MW. We also find that these GCs are concentrated in both the IOM and action space diagrams, and occupy unique regions. We select GSE-a as follows
\begin{equation}
\begin{aligned}
    & ~ (-2\times10^5<E<-1.4\times10^5~\rm{km^2\,s^{-2}})\\
    &\land ~ (-800<L_z<0~\rm{kpc\,km\,s^{-1}})\\
    &\land ~ (-0.4<J_\phi/J_{\rm tot}<-0.1)\\
    &\land ~ ((J_z-J_R)/J_{\rm tot}<-0.5)\\
    &\land ~ (z_{\rm max}<2~\rm{kpc})\\
\end{aligned}
\label{eq:gea}
\end{equation}
In addition, their high ecc ($>0.7$) makes them easily distinguishable from disk GCs. Their energy and Apo are also well above the limits of the bulge GCs.

Based on the orbital features and the action space of GSE-a GCs, we think that NGC 6121 is more likely a member of the GSE-a GCs. \citet{Duncan2020} considered NGC 6121 to be an in situ GC according to its AMR. However, due to the large error in the AMR estimation, we think that NGC 6121 is more likely to belong to GSE-a. In M19 and BB21, NGC 6121 is taken as the L-E (low energy) GC and it is a Kraken GC in C22 (See Table~\ref{tab:duibi}). As shown in Section~\ref{sec:kraken}, Kraken GC is the L-E GC.  From Figure~\ref{fig:space}, we see that the GSE-a and the Kraken GCs have clear difference in the action space. Therefore, we propose that NGC 6121 is a GSE-a GC.



For GC VVV CL001, although its orbital properties (ecc, Apo and Peri) are similar to a GSE-a GC, its orbit
is obviously retrograde and it cannot be related to other progenitors.  This makes VVV CL001 the lowest energy  Pot accretion GC identified in this paper.

Although we have not found relevant counterparts to GSE-a in other existing studies, we think that GSE-a is an accretion event distinct from GSE.

\subsubsection{GSE-b}
\label{sec:Two}


Shown as orange upward-pointing  triangles in Figs.~\ref{fig:fehago}--\ref{fig:space}, 
we find that GSE-b GCs have significantly lower ecc than that of the GSE GCs.
Moreover, the GSE-b GCs in the action space map  differ significantly from the  GSE and Seq GCs \citep[][]{myeong2019}. The definition of the GSE-b GCs is
\begin{equation}
\begin{aligned}
    &(-1.8\times10^5<E<-1\times10^5~\rm{km^2\,s^{-2}})\\
    &\land ~ (0<L_z<1200~\rm{kpc\,km\,s^{-1}})\\
    &\land ~ (0.1<J_\phi/J_{\rm tot}<0.5)\\
    &\land ~ (0<(J_z-J_R)/J_{\rm tot}<0.4)\\
    &\land ~ (0.5<\mathrm{ecc}<0.7)
\end{aligned}
\label{eq:geb}
\end{equation}
Although GSE-b GCs are close to the GSE GCs in IOM, and their orbital ecc is close to Seq,  their action space positioning distinguishes them. In addition, it seems that the IOM and action space range of the GSE-b GCs are similar to those of Thamnos in \citet{naidu2020} \footnote{In order to compare our result with the Thamnos, we also used the same potential in \citet{naidu2020}, and we find that the difference of GSE-b GCs in the action space from our potential and the potential in  \citet{naidu2020} is tiny.} 
and Pontus in \citep{malhan2022a, malhan2022} and the IOM range of GSE-b slightly different from that of the Thamnos in \citet{koppelman2019}. 

The GSE-b GCs have relatively larger values of $J_z$ than GSE and Seq GCs. In addition, the AMR of GSE-b GCs is slightly steeper than that of GSE GCs.

Finally, we believe that GSE-b, representing a retrograde system with average energy lower than Seq, might possibly be associated with Thamnos and Pontus. Hence, we assess GSE-b as probably an accretion event in the MW's history.

\subsubsection{GSE-c}
\label{sec:Three}
GSE-c GCs are displayed as brown squares in Figs.~\ref{fig:fehago}--\ref{fig:space}. An open question is whether to classify GSE as a prograde or a retrograde accretion. In the past, it was thought that GSE may be a slight retrograde accretion to the MW, but the trend is not clear \citep{Massari2019, myeong2019}.  Recently, \citet{Kwang_Kim_2021} found that GSE tends to be prograde in the inner halo and retrograde in the outer halo. They pointed out that this may have been caused by more than two different accretions. After identifying GSE-a and GSE-b GCs, we find that the remaining GCs manifest clear prograde and retrograde rotations in IOM and action space. We take GCs with the prograde rotation as GSE-c.  We show GSE-c as brown squares in Figs.~\ref{fig:fehago}--\ref{fig:space}.
The remaining part of the GSE GCs then shows clear retrograde rotation. The GSE-c GCs are defined as 

\begin{equation}
\begin{aligned}
    & ~ (-1.8\times10^5<E<-1\times10^5~\rm{km^2\,s^{-2}})\\
    &\land ~ (-1000<L_z<0~\rm{kpc\,km\,s^{-1}})\\
    &\land ~ (-0.5<J_\phi/J_{\rm tot}<-0.1)\\
    &\land ~ (-0.5<(J_z-J_R)/J_{\rm tot}<0)\\
\end{aligned}
\label{eq:gec}
\end{equation}
From these selection criteria, we assign NGC 3333 (M9), NGC 6584 and Rup 106 to GSE-c.

Comparing with \citet{naidu2020}, we find that GSE-c overlaps with Wukong in IOM, but there is some deviation in action space. We think that GSE-c might be associated with Wukong and be part of the dwarf galaxy accretion in the MW's history.

\subsubsection{Kraken}
\label{sec:kraken}

Shown as sky blue downward-pointing triangles in Figs.~\ref{fig:fehago}--\ref{fig:space}, the Kraken progenitor was first proposed by \citet{kruijssen2018}, and corresponds to the low energy group in \citet{Massari2019}. The Kraken GCs have orbits similar to GSE and bulge GCs. Kraken GCs have a more diffuse distribution in action space. Due to their closeness to the IOM of bulge GCs, it is difficult to distinguish Kraken GCs from the bulge GCs. While for individual GCs, age can help us distinguish them \citep{2022arXiv}. The selection criteria for Kraken GCs are
\begin{equation}
\begin{aligned}
    &(-2.1\times10^5<E<-1.7\times10^5~\rm{km^2\,s^{-2}})\\
    &\land ~ (-400<L_z<100~\rm{kpc\,km\,s^{-1}})\\
\end{aligned}
\label{eq:kraken}
\end{equation}
Compared with the M19 low-energy sample, the Kraken sample does not contain NGC 6093 (M80), NGC 6121 (M4), NGC 6256, NGC 6441, NGC 6544 and NGC 6681 (M70).

\subsubsection{Potential Associate (Pot)}
\label{sec:he}


These Potential (Pot) GCs are those that cannot be associated with known major merger events. Therefore, we think the Pot GCs may be from small accretion events. Some previous studies have attempted to detect such small-scale accretion events \citep[][]{2018MNRAS.475.1537M, myeong2018, koppelman2019, necib2020, Fiorentin_2021, Horta2020}. 
Although most Pot GCs have high energy, some small accretion events may occur early in the MW's lifetime in low energy regions. It is difficult, however, to distinguish them from other GCs.

\section{MW mass estimation method}
\label{sec:method}


We now consider how we can use the GCs to estimate the mass of the MW.  We describe our estimation method, including the mass estimator, the potential model and how we fit the parameters needed for the estimator from the data. 


The spherical Jeans equation has been widely applied to estimate the mass enclosed within a certain radius. Based on this equation, \citet{Watkins2010} introduced a family of traced mass estimators which utilizes the discrete positional and kinematical data of different tracers.
In this investigation, we apply an estimator from \citet{Watkins2010} employing distance and total velocity,
\begin{equation}
    M(<r_{\rm max})=\frac{\alpha+\gamma-2\beta}{G(3-2\beta)}r^{1-\alpha}_{\rm max} \langle v^2r^{\alpha}\rangle
	\label{eq:quadratic}
\end{equation}
where $r_{\rm max}$ is the most distant GC's Galactocentric distance, $v$ is the total velocity of the GC, $\alpha$ and $\gamma$ are the power law indices for the underlying gravitational potential $(\Phi\propto\ r^{-\alpha})$ and the number density profile $(\rho\propto\ r^{-\gamma})$ respectively, and $\beta$ is the velocity anisotropy parameter of the GC sample.

\subsection{Anisotropy}
\label{sec:anisotropy}

The velocity dispersion anisotropy parameter $\beta$ is defined as \citep{binney2011galactic} 
\begin{equation}
    \beta=1-\frac{\sigma^2_{\theta}+\sigma^2_{\phi}}{2\sigma^2_r}
    =1-\frac{\langle v^2_{\theta}\rangle-\langle v_{\theta}\rangle^2+\langle v^2_{\phi}\rangle-\langle v_{\phi}\rangle^2}{2(\langle v^2_r\rangle-\langle v_r\rangle^2)}
	\label{eq:anisotropy}
\end{equation}
where $\sigma_{\theta}$ and $\sigma_{\phi}$ are velocity dispersions along the two tangential directions, and $\sigma_r$ is the radial velocity dispersion. Purely circular orbits correspond to $\beta=-\infty$, radial orbits correspond to $\beta=1$ and for an isotropic distribution of orbits $\beta=0$. Note that the velocity here is given in spherical coordinates, while in Section~\ref{sec:data} the velocity is expressed in Cartesian coordinates. 

\begin{figure}
\includegraphics[width= \columnwidth]{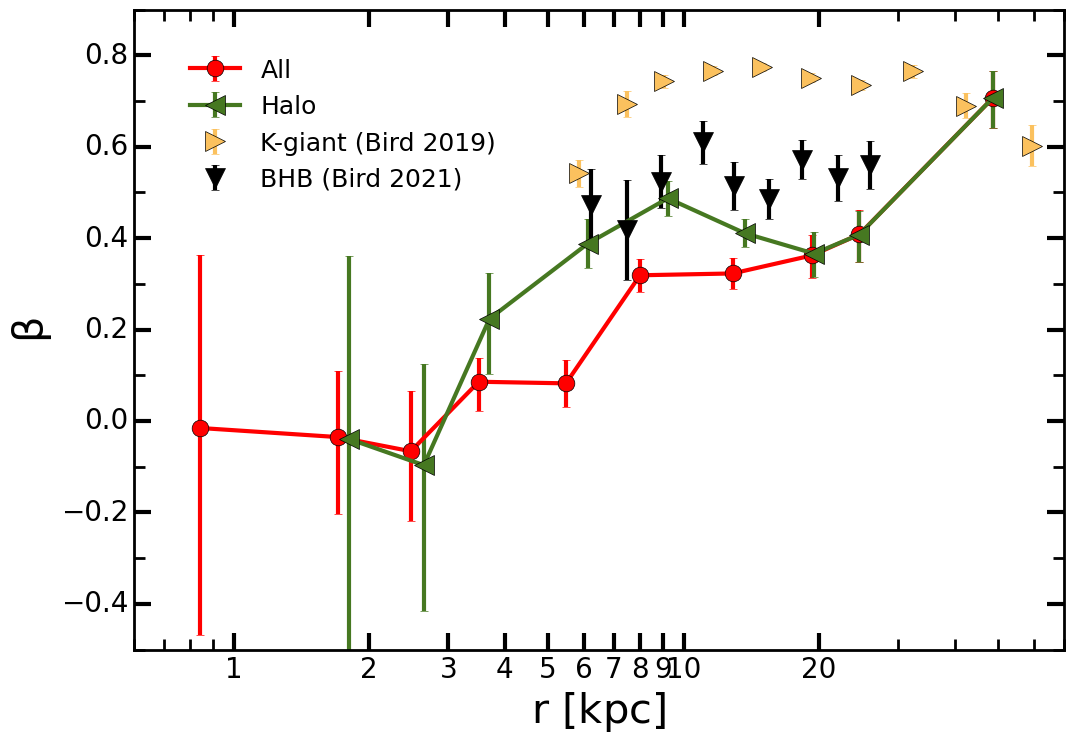}
\caption{Distribution of the anisotropy parameter $\beta$ for different tracers along the Galactocentric radius. We used the jackknife method to estimate the statistical error. The total uncertainty is the sum of the observational and statistical uncertainties. The `All' sample includes all GCs, while the `Halo' sample excludes bulge and disk GCs. The K-giant and BHB data from \citet{Bird2019} and \citet{bird2021} are also displayed.}
    \label{fig:beta}
\end{figure}

As shown in Fig.~\ref{fig:beta}, we calculate the $\beta$ value at different Galactocentric radii. We divide the 0.69 to 60 kpc range into 10 logarithmically uniform shells. Each shell contains approximately 15 GCs for the All sample. Due to the small number of GCs at larger radii, we only bin GCs at less than 60\,kpc. In Fig.~\ref{fig:beta}, the $\beta$ values of each bin are calculated with data in three adjacent bins except the first and last bins. 
The mean values of the radius are taken as the radius corresponding to $\beta$ in those bins. Therefore, 2/3 of small bins overlap for two close radii in the major diagram. The aim of the overlapping bins is to make the curve as smooth as possible to reflect the general trend of changes in $\beta$, and to reduce Poisson noise.
Non-overlapping bins would lead to large Poisson error and change the $\beta$ curve drastically. Although the average $\beta$ value in each bin depends somewhat on the particular gridding, the overall trend is robust.
Our result indicates that the drop of $\beta$ near 20\,kpc may be related to Sgr \citep[][]{Bird2019, bird2021}. The other drop at $\sim$3-7 kpc could be the result of the Seq dwarf galaxy.

In Fig.~\ref{fig:beta}, we show the $\beta$ value for the ``All" and ``Halo" samples. The ``All" sample includes all GCs considered in this paper, while the ``Halo" sample excludes the bulge and disk GCs. The GC numbers for each sample are given in Table~\ref{tab:mass}.
It is seen that $\beta$ increases with the Galactocentric radius, which means that the orbits of more distant GCs are more radial. Moreover, the $\beta$ curves of `All' and `Halo' are similar, with the only difference being in the middle radii range 4-15 kpc.


Compared with related studies using different tracers, our results around 10\,kpc are significantly lower than RR Lyrae type variables, $\beta\sim0.8$ \citep{Hattori2020}, and K giants, $\beta\sim0.75$ \citep{Bird2019}, but only slightly below ($\sim 0.1$) that of blue horizontal branch (BHB) stars, $\beta\sim0.5$ \citep{utkin2020, bird2021}, which are also ancient tracers.  The trend of the `Halo' curve is consistent with that in \cite{bird2021}; the only difference is that our $\beta$ values are slightly smaller than theirs in the overlapping radius regions. The GC sample allows us to estimate $\beta$ close to the central region of the MW, where $\beta$ is nearly isotropic and slightly smaller than 0.

As shown in \cite{bird2021}, the difference in $\beta$ values from the K giants and BHB is likely due to the different metallicity distributions. 
We have checked that if we only use the GCs  with $\rm [Fe/H]<-1$, the $\beta$ distribution from our sample is fully consistent with that in   \cite{bird2021} by using the BHB sample with the similar metallicity selection.


\subsection{Number density}
\label{sec:number density}

Assuming the GC number density takes the form $\rho = k r^{-\gamma}$, where $k$ is a constant, the GC count in a given spherical shell is given by
\begin{equation}
    N(r)=\int_{r_{\rm min}}^{r_{\rm max}}{4\pi r^2\rho dr}=4\pi k\frac{r^{3-\gamma}_{\rm max}-r^{3-\gamma}_{\rm min}}{3-\gamma}.
	\label{eq:numberdensity}
\end{equation}
Therefore,
\begin{equation}
    k = \frac{(3-\gamma)N_{\rm tot}}{4\pi(r_{\rm max}^{3-\gamma}-r_{\rm min}^{3-\gamma} )},
	\label{eq:kkk}
\end{equation}
where $N_{\rm tot}$ is the total number of GCs and $\gamma \neq 3$.
The  power law index $\gamma$ of the GC number density profile is obtained by using all the GCs. Two parameters $\rm \gamma_{\rm inner}$ and $\rm \gamma_{\rm outer}$ are defined as the indices in the inner and outer regions, respectively. We fit a broken power law to the data, and the posterior probability distribution $p(\gamma, \theta|N(r))=p(\gamma|N(r))$ is the probability of model parameters $\gamma$, conditional on a set of data $N(r)$:
\begin{equation}
\begin{split}
    p(\gamma|N(r))&\propto p(N(r)|\theta)p(\gamma, \theta)\\
    &\propto \mathrm{Poisson}(N(r),\theta)p(\theta|\gamma)
	\label{eq:beyesi}
\end{split}
\end{equation}
where $\theta$ is possible values, and $p(N(r)|\theta)$ is the likelihood assuming a Poisson distribution. $p(\gamma, \theta) = p(\theta|\gamma)p(\gamma)$ is the joint prior distribution on $\theta$ and $\gamma$. We assume that $p(\gamma)$ is uniformly distributed.

\begin{figure}
	\includegraphics[width= \columnwidth]{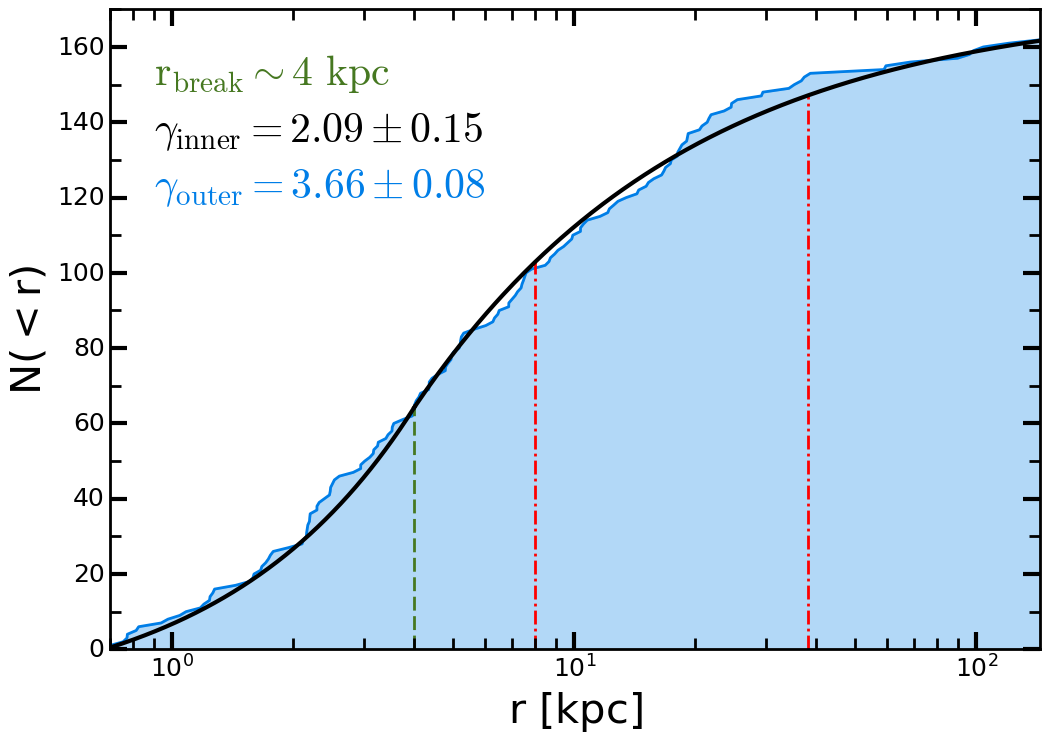}
    \caption{The cumulative number profile for 162 MW GCs. (To ensure the completeness of the data as much as possible, we use all of the GCs presented in \citet{Baumgardt_2021}, and so there is a slight difference in the number of GCs.) A broken power law model with an index $\gamma_{\rm inner}$ in the inner regions and $\gamma_{\rm outer}$ in the outer regions is adopted. The blue cumulative histogram shows the data,  while the solid black line is the best-fitting power law curve. The best-fitting parameters are written in the top-left corner of the plot. The uncertainty in this value is the uncertainty in the Bayesian method. The green vertical dashed line shows the broken radius at $\sim\ 4$\,kpc. Two red vertical dashed lines represent the GCs in the best sample at the innermost radius 8 kpc and outermost radius 37.3 kpc (see Sec~\ref{sec:potential}).
}
\label{fig:gamma}
\end{figure}

Fig.~\ref{fig:gamma} shows in blue the cumulative number density distribution of all MW GCs.
The black solid line is the maximum likelihood broken power law of the data, and the dashed black line represents the broken radius at which the power law index changes. Here we take broken radius $r_{\rm break}\sim 4$ kpc to distinguish the bar region and other regions \cite{Wang12,Wang13}.  
The parameter fit values with statistical uncertainties are expressed in the upper left corner. Since we focus on the halo region of the MW, we adopt $\gamma=3.66\pm0.08$ for our mass analysis. 

\subsection{Gravitational potential}
\label{sec:potential}

\begin{figure}

	\includegraphics[width= \columnwidth]{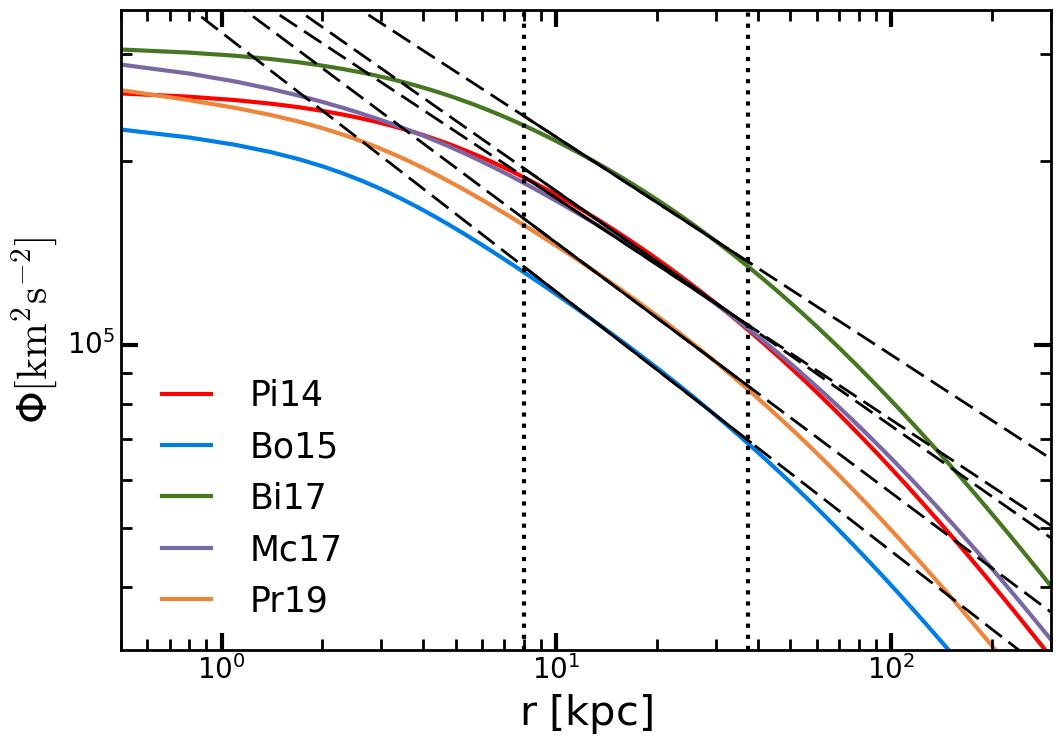}
    \caption{Potential distribution along the radius for different models. Two vertical black dotted lines represent the best sample range (see Sec~\ref{sec:mass of the mw}). The black solid lines show the region of the best-fitting power laws, and the dashed lines signify the extended region outside the best-fitting power laws region. Since there is no clear difference between halo and total potentials at large $r$, we use the total potential here. }
    \label{fig:figure6}
\end{figure}

In order to find a suitable  $\alpha$ value, we consider five general MW potential models. They are Pi14, Bo15, Bi17, Mc17, and Pr19.
We estimate the gradient of the underlying potential, and summarize the values of the  $\alpha$ parameter for the potential models in Fig.~\ref{fig:figure6}. We fit the curves in the best radius range by linear regression, and use the average $\alpha$ of the five models as the final $\alpha$ we use. In the best fit range, we obtain $\alpha=0.388\pm0.049$.

\section{The mass of the MW}
\label{sec:mass of the mw}

In using the spherical Jeans equation to derive the MW mass, it is assumed that a dynamical equilibrium state is reached, and the tracers follow the same distribution  as the particles in the equilibrium system. The GCs from recent accretion events may not completely satisfy this condition, which may introduce error in the estimated velocity dispersion anisotropy $\beta$ and the MW mass. To quantify this error, and to obtain a better estimate of the mass, 
below we try to assess this deviation from the data themselfes, and then to select a sample which is less affected by such deviations.

\begin{figure}

\includegraphics[width= \columnwidth]{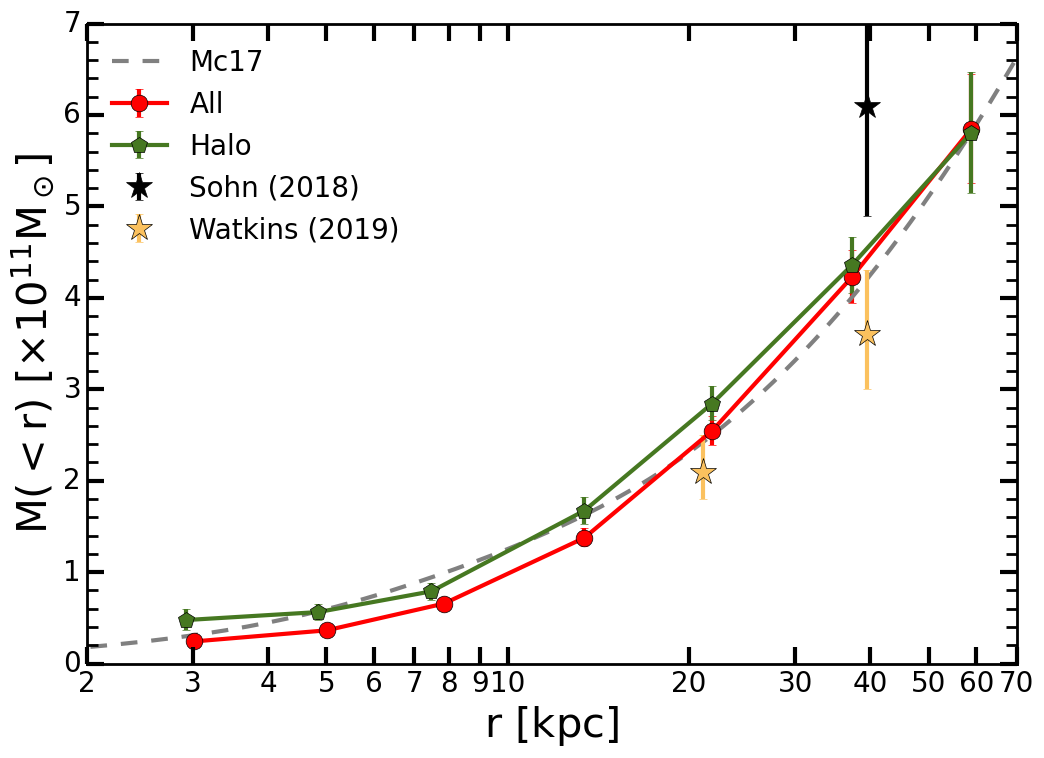}
    \caption{Variation of the MW mass with Galactic radius. The red and green lines represent the results from All and Halo sample, respectively. The gray dashed line shows the mass profile given by Mc17. The black star is the mass calculated by using the same GCs as in \citet{Sohn2018}. The yellow stars are the result by using the GCs in  \citet{Watkins2019}.  }
    \label{fig:mass}
\end{figure}

Once we have the values of the potential power law index $\alpha$, the anisotropy parameter $\beta$, and the density power law index $\gamma$, we can use Equation~\ref{eq:quadratic} to estimate the MW mass enclosed within $r_{\rm max}$. In this equation, $\alpha$, $\beta$, and $\gamma$ are usually assumed to be constants. However, it is known that these parameters actually vary with the radius \citep[e.g.][]{2012ApJ...761...98K,Watkins2019}. Therefore, it is necessary to select an appropriate radius range. We take $8.0$\,kpc$\,<r<37.3$\,kpc as the range of radius for the following reasons: 
\begin{enumerate}
 \item By taking $r> 8 \kpc$ we exclude all bulge GCs and most of the disk GCs, so that most of the  remaining GCs  come from the halo;
 \item The parameters of $\alpha$, $\beta$, and $\gamma$ in the selected range are assumed to be constants  \citep{Watkins2019}, so we need to take a range in which these parameters are nearly constant;
 \item Outside of this range, the number density of the GC sample is too small. 
\end{enumerate}

The estimated mass  of the MW is $M(<37.3 \kpc)=0.423_{-0.021}^{+0.022}\times10^{12}M_{\odot}$ from the  `All' sample, \footnote{In order to keep consistency with Section~\ref{sec:anisotropy}, here we still use the name `All' sample and `Halo' sample, but please keep in mind that both of them are constrained by $8.0$\,kpc$\,<r<37.3$\,kpc.} and $M(<37.3 \kpc)=0.437_{-0.022}^{+0.022}\times10^{12}M_{\sun}$ from the `Halo' sample.  
Although our analysis concentrates on the radius range from 8 kpc to 37.3 kpc, it can be extended to smaller or larger radii.      

Fig.~\ref{fig:mass} features the mass distribution along the MW radius for the different selected samples. It is seen that the estimated mass from the `All' GC sample is slightly smaller than that obtained from the `Halo' sample, which indicates that the bulge and disk GCs in this case tend to decrease the estimate of the MW mass slightly. It is also noted that the mass profile from our Halo sample is consistent with that in Mc17 by considering the measurement error of the mass.   

In the same figure, we also show the results from \cite{Sohn2018} and \cite{Watkins2019}. These two results are also obtained by using the GC samples. It is seen that the mass estimated in \cite{Sohn2018} is larger than $20\%$ of the mass from our Halo sample. The error bar of the mass is large in their results due to a small number of GCs in their sample (only 16 GCs).  At r$=21.1$ kpc, and $r=39.5$ kpc, our estimated masses from both All and Halo samples are larger than those in \cite{Watkins2019}.

In order to estimate the virial mass of the MW, we assume that the MW has a similar potential as that in Mc17, the only difference is that we assume the dark matter halo follows the NFW profile \citep{nfw, nfw1997}.  We take $M_{200}$ as the virial mass, and it is defined as the total mass enclosed within a radius $r_{200}$, inside which the mean density is 200 times the critical density of the universe. The estimated $M_{200}$ values are $1.11_{-0.18}^{+0.25}\times10^{12}M_{\odot}$  and $1.16_{-0.18}^{+0.25}\times10^{12}M_{\odot}$
for the All sample and Halo sample, respectively (See Table~\ref{tab:mass}). As noted in \cite{Wang2020},  \cite{Sohn2018} and \cite{Watkins2019} give values of $1.71_{-0.79}^{+0.97}\times10^{12}M_{\odot}$  and $1.29_{-0.44}^{+0.75}\times10^{12}M_{\odot}$, respectively.  Our $M_{200}$ values are smaller,  but the results are still consistent with each other within the error bar. Our results are closer to the estimates of \citet{Callingham2019} based on the dynamics of MW satellites, and those of \cite{zhai18} based on K giants, which are   $M_{200}=1.17_{-0.15}^{+0.21}\times10^{12}M_{\sun}$, and $M_{200}= 1.11_{-0.20}^{+0.24}\times10^{12}M_{\odot}$, respectively.

\section{summary and discussion}
\label{sec:conclusions}

\begin{table}
	\centering
	\caption{GC sample, anisotropy parameter, Mass within 8-37.3 kpc, $M_{200}$ and number of GCs in the corresponding sample.
}
	\label{tab:mass}
	\begin{tabular}{lcccr} 
		\hline
		Sample & $\beta$ & $M(<37.3 \kpc)$ & $M_{200}$ & Number\\
		 &  & $[\times10^{12}M_{\sun}]$ & $[\times10^{12}M_{\sun}]$\\
		\hline
		
		All & $0.315_{-0.049}^{+0.055}$ & $0.423_{-0.021}^{+0.022}$ & $1.11_{-0.18}^{+0.25}$ & 49\\
		Halo & $0.351_{-0.055}^{+0.058} $& $0.437_{-0.022}^{+0.022}$ & $1.16_{-0.18}^{+0.25}$ & 46\\
		
		\hline
	\end{tabular}
\end{table}

With the available high-precision PMs from the Gaia EDR3 \citep{Vasiliev2021}, we have the full six-dimensional phase space data for 159 GCs in the MW. We have revised the classification for these GCs by combining the AMR, IOM, action space and orbit information. The major changes in the classification have been the introduction of multi-faceted membership criteria to the various subgroups,
and splitting the GSE subgroup into smaller groupings. Moreover, we have measured the anisotropy parameter $\beta$ and the mass of the MW. The main results of this paper can be summarized as follows.    

\begin{enumerate}
\item
We found that 88 GCs may have been accreted  by the MW, and we assessed their possible associations with the progenitor of five known merger events: the Sagittarius dwarf, the Sequoia galaxy, Helmi Streams, GSE, and Kraken. We isolated three new structures (GSE-a, GSE-b and GSE-c) from the GSE. These GCs are likely unrelated to the GSE, but further certification is required.
\item
There are 26 GCs that could not be associated with the known progenitors, and these members may have come from multiple small accretion events in the MW's history. The detailed classification results of the GCs are given in Table~\ref{tab:A1}.
\item
The anisotropy parameter $\beta$ depends strongly on the selected GC sample. The $\beta$ values increase with the Galactic radius. The $\beta$ values from our Halo GC sample are smaller than those derived from the BHB stars in \cite{bird2021}. 
However, if we only use the GCs with $\rm [Fe/H]<-1$, our $\beta$ distribution will be totally consistent with that in  \cite{bird2021}, which again shows that the $\beta$ distribution depends on the metallicity of the sample. The average $\beta$ values for different GC samples are expressed in Table~\ref{tab:mass}. Compared with the BHB and K giants, the GC sample allows us to determine the $\beta$ distribution close to the center of the MW. The $\beta$ is negative and close to zero in the central region of the MW, which indicates that the GC orbits prefer to be circular in this region. 
\item
We use the scale-free mass estimator of \citet{Watkins2010} to estimate the MW mass, and obtain
$M(<37.3 \kpc)=0.423_{-0.021}^{+0.022}\times10^{12}M_{\sun}$ and $M_{200}=1.11_{-0.18}^{+0.25}\times10^{12}M_{\sun}$ from the Halo sample. Compared with the previous studies, our results are similar with $M_{200}=1.17_{-0.15}^{+0.21}\times10^{12}M_{\sun}$ as obtained by \citet{Callingham2019} and $M_{200}= 1.11_{-0.20}^{+0.24}\times10^{12}M_{\odot}$  gotten by \cite{zhai18}. 

\end{enumerate}

The mass measurement method adopted in this paper is the same as in  \citet{Sohn2018} and \citet{Watkins2019}. The MW mass obtained in  \citet{Sohn2018} is  $M_{200}=1.71_{-0.79}^{+0.97}\times10^{12}M_{\sun}$ and in \citet{Watkins2019} is $M_{200}=1.29_{-0.44}^{+0.75}\times10^{12}M_{\sun}$. By carefully comparing our differences, we found that our most prominent difference is the selected sample.  \citet{Eadie2019} estimated mass $M_{200}=0.77_{-0.24}^{+0.46}\times10^{12}M_{\sun}$,  and obtained $M(r<39.5\ \kpc)=0.33_{-0.07}^{+0.12}\times10^{12}M_{\sun}$. Our result is similar to theirs.  \citet{Eadie2019} also noted that there is a major difference between their result and the result of \citet{Vasiliev2019}, but they did not find the exact reason for this difference. From our studies, we can see that their difference is most likely due to the different data samples.


As more low-luminosity GCs are discovered \citep[]{2017ApJ...849L..24M, 2019MNRAS.484L..90C, 2021BAAA...62..107M, 2021A&A...654A..39O, 2022A&A...659A.155G}, and with the improvement of the astrometry accuracy by the Gaia mission, more six-dimensional observations of GCs will become available, which will refine the mass estimate of the MW.

\section*{Acknowledgements}

We thank the referees for the constructive and detailed comments for improving the paper. The title about GCs was originally inspired by Duo Li. We thank Furen Deng for helpful discussions and suggestions. We also thank Sarah A. Bird for providing us the beta values from her paper. 
We acknowledge the support by the National Key R\&D Program (No. 2017YFA0402603) and the Inter-government cooperation Flagship program (grant No.~2018YFE0120800), the  National Natural Science Foundation of China (NSFC, Grant Nos. 11773034 and 11633004), the Chinese Academy of Sciences (CAS) Strategic Priority Research Program XDA15020200 and and the CAS Interdisciplinary Innovation Team (JCTD- 2019-05).



\bibliographystyle{raa}
\bibliography{ms} 




\appendix
\section{The Data Table}

\onecolumn
\small
\renewcommand{\arraystretch}{1}
\setlength\tabcolsep{3pt}
\setlength{\LTcapwidth}{\textwidth}
\begin{longtable}[c]{l|c|c|c|c|c|c|c|c|r}

\caption{Some Results and Observations Used in this Paper. Column (1): the name of GCs. Column (2): Associated progenitors in this paper. Column (3): right ascension. Column (4): declination. Column (5): metallicity. Column (6): distance from Sun. Column (7): velocity of the line-of-sight. Column (8): proper motion in right ascension. Column (9): proper motion in declination. Column (10): correlation coefficient between $\mu_\alpha$ and $\mu_\delta$. Columns (3)-(10) are the observations we used, the specific data sources are described in Section~\ref{sec:data}.}

\label{tab:A1}\\

\hline
Cluster & Type & $\alpha$ & $\delta$ & Fe/H & $d$ & $r_{\rm LOS}$ & $\mu_\alpha$ & $\mu_\delta$ & $\mathrm{cor}r_\mu$\\
& & (deg) & (deg) & & (kpc) & (km $s^{-1}$) & (mas $yr^{-1}$) & (mas $yr^{-1}$) & \\
\hline
\endfirsthead
\hline
Cluster & Type & $\alpha$ & $\delta$ & Fe/H & $d$ & $r_{\rm LOS}$ & $\mu_\alpha$ & $\mu_\delta$ & $\mathrm{corr}_\mu$\\
& & (deg) & (deg) & & (kpc) & (km $s^{-1}$) & (mas $\mathrm{yr}^{-1}$) & (mas $\mathrm{yr}^{-1}$) & \\
\hline
\endhead 

\hline
\endfoot 

NGC 104 (47 Tuc) & D& 6.024 & -72.081 & -0.72 & 4.52 $\pm\,$ 0.03 & -17.45 $\pm\,$ 0.16 & 5.252 $\pm\,$ 0.021 & -2.551 $\pm\,$ 0.021 & 0.00 \\ 
NGC 288 & GSE-b& 13.188 & -26.583 & -1.32 & 8.99 $\pm\,$ 0.09 & -44.45 $\pm\,$ 0.13 & 4.164 $\pm\,$ 0.024 & -5.705 $\pm\,$ 0.024 & 0.01 \\ 
NGC 362 & GSE& 15.809 & -70.849 & -1.26 & 8.83 $\pm\,$ 0.1 & 223.12 $\pm\,$ 0.28 & 6.694 $\pm\,$ 0.025 & -2.535 $\pm\,$ 0.024 & 0.00 \\ 
Whiting 1 & Sgr& 30.737 & -3.253 & -0.7 & 30.59 $\pm\,$ 1.17 & -130.41 $\pm\,$ 1.79 & -0.228 $\pm\,$ 0.065 & -2.046 $\pm\,$ 0.056 & 0.03 \\ 
NGC 1261 & GSE& 48.068 & -55.216 & -1.27 & 16.4 $\pm\,$ 0.19 & 71.34 $\pm\,$ 0.21 & 1.596 $\pm\,$ 0.025 & -2.064 $\pm\,$ 0.025 & 0.01 \\ 
Pal 1 & Pot& 53.334 & 79.581 & -0.65 & 11.18 $\pm\,$ 0.32 & -75.72 $\pm\,$ 0.29 & -0.252 $\pm\,$ 0.034 & 0.007 $\pm\,$ 0.037 & 0.08 \\ 
AM 1 (E 1) & Pot& 58.760 & -49.615 & -1.7 & 118.91 $\pm\,$ 3.4 & 118 $\pm\,$ 14.14 & 0.291 $\pm\,$ 0.107 & -0.177 $\pm\,$ 0.086 & -0.22 \\ 
Eridanus & Pot& 66.186 & -21.187 & -1.43 & 84.68 $\pm\,$ 2.89 & -23.15 $\pm\,$ 0.73 & 0.51 $\pm\,$ 0.039 & -0.301 $\pm\,$ 0.041 & -0.09 \\
\end{longtable}

\renewcommand{\arraystretch}{1}
\setlength\tabcolsep{0pt}
\setlength{\LTcapwidth}{\textwidth}
\setlength\LTleft{-0.2in}

\begin{longtable}{l|c|c|c|c|c|c|c|c}
\caption{Positions, Velocities, Apocenter and Pericenter in Galactocentric Coordinate System for the full 159 GC Sample. Columns (1): the name of GCs. Columns (2)-(4): Position in the $x$, $y$ and $z$ directions. Columns (5)-(7):  Velocity components. Columns (8)-(9): apocenter and pericenter.}

\label{tab:A2}\\

\hline
Cluster & X & Y & Z &  U & V & W & Apo &  Peri\\
& (kpc) & (kpc) & (kpc) & (km $s^{-1}$) & (km $s^{-1}$) & (km $s^{-1}$) & (kpc) & (kpc)\\
\hline
\endfirsthead
\hline
Cluster & X & Y & Z &  U &  V &  W & Apo &  Peri\\
& (kpc) & (kpc) & (kpc) & (km $s^{-1}$) & (km $s^{-1}$) & (km $s^{-1}$) & (kpc) & (kpc)\\
\hline
\endhead 

\hline
\endfoot 

NGC 104 (47 Tuc)& -6.31 $\pm\,$ 0.04 & -2.59 $\pm\,$ 0.02 & -3.17 $\pm\,$ 0.02 & -77.30 $\pm\,$ 1.87 & 163.86 $\pm\,$ 5.42 & 46.17 $\pm\,$ 0.98 & 7.57 $\pm\,$ 0.03 & 5.59 $\pm\,$ 0.24 \\ 
NGC 288& -8.29 $\pm\,$ 0.03 & 0.05 $\pm\,$ 0.00 & -8.97 $\pm\,$ 0.09 & -4.49 $\pm\,$ 2.05 & -55.91 $\pm\,$ 6.36 & 50.39 $\pm\,$ 0.88 & 12.38 $\pm\,$ 0.08 & 2.03 $\pm\,$ 0.24 \\ 
NGC 362& -5.00 $\pm\,$ 0.05 & -5.21 $\pm\,$ 0.06 & -6.37 $\pm\,$ 0.07 & -93.77 $\pm\,$ 2.91 & -105.71 $\pm\,$ 6.05 & -67.58 $\pm\,$ 1.50 & 12.36 $\pm\,$ 0.22 & 0.26 $\pm\,$ 0.12 \\ 
Whiting 1& -22.51 $\pm\,$ 0.55 & 4.74 $\pm\,$ 0.18 & -26.65 $\pm\,$ 1.02 & 247.22 $\pm\,$ 10.62 & 20.93 $\pm\,$ 12.80 & -9.72 $\pm\,$ 6.76 & 61.05 $\pm\,$ 5.69 & 21.33 $\pm\,$ 1.89 \\ 
NGC 1261& -8.12 $\pm\,$ 0.04 & -10.07 $\pm\,$ 0.12 & -12.93 $\pm\,$ 0.15 & 79.17 $\pm\,$ 2.79 & 50.55 $\pm\,$ 5.81 & 68.60 $\pm\,$ 1.98 & 20.93 $\pm\,$ 0.26 & 1.27 $\pm\,$ 0.23 \\ 
Pal 1& -14.97 $\pm\,$ 0.20 & 8.08 $\pm\,$ 0.24 & 3.68 $\pm\,$ 0.11 & 63.70 $\pm\,$ 2.28 & 199.19 $\pm\,$ 5.63 & -24.97 $\pm\,$ 2.07 & 18.52 $\pm\,$ 0.54 & 14.40 $\pm\,$ 0.54 \\ 
AM 1 (E 1)& -24.31 $\pm\,$ 0.46 & -77.22 $\pm\,$ 2.18 & -88.96 $\pm\,$ 2.51 & 18.44 $\pm\,$ 45.84 & 20.91 $\pm\,$ 47.97 & 43.25 $\pm\,$ 42.33 & 125.95 $\pm\,$ 7.24 & 19.21 $\pm\,$ 18.44 \\ 
Eridanus& -58.33 $\pm\,$ 1.71 & -39.22 $\pm\,$ 1.34 & -55.74 $\pm\,$ 1.91 & 53.16 $\pm\,$ 13.19 & 51.80 $\pm\,$ 17.46 & 140.21 $\pm\,$ 12.47 & 124.00 $\pm\,$ 5.58 & 13.88 $\pm\,$ 4.37 \\
\end{longtable}


\end{document}